\pdfoutput=1
\documentclass[aps,prl,twocolumn,superscriptaddress]{revtex4-2}

\usepackage{graphicx,xcolor}
\usepackage{amsmath,amssymb,amsthm}
\usepackage[pdfencoding=auto]{hyperref}
\usepackage{physics}
\usepackage[small,bf]{caption}
\usepackage[subrefformat=parens, labelformat=simple,singlelinecheck=false]{subcaption}
\usepackage{siunitx}
\hypersetup{colorlinks=true,allcolors=black}

\begin{document}

\title{Linear Optical Quantum Computation with Frequency-Comb Qubits and Passive Devices}

\author{Tomohiro Yamazaki}
\email{tomohiro.yamazaki@ntt.com}
\affiliation{Graduate School of Engineering Science, Osaka University, Toyonaka, Osaka 560-8531, Japan}
\affiliation{Center for Quantum Information and Quantum Biology, Osaka University, Toyonaka, Osaka 560-0043, Japan}

\author{Tomoaki Arizono}
\affiliation{Graduate School of Engineering Science, Osaka University, Toyonaka, Osaka 560-8531, Japan}

\author{Toshiki Kobayashi}
\affiliation{Graduate School of Engineering Science, Osaka University, Toyonaka, Osaka 560-8531, Japan}
\affiliation{Center for Quantum Information and Quantum Biology, Osaka University, Toyonaka, Osaka 560-0043, Japan}

\author{Rikizo Ikuta}
\affiliation{Graduate School of Engineering Science, Osaka University, Toyonaka, Osaka 560-8531, Japan}
\affiliation{Center for Quantum Information and Quantum Biology, Osaka University, Toyonaka, Osaka 560-0043, Japan}

\author{Takashi Yamamoto}
\affiliation{Graduate School of Engineering Science, Osaka University, Toyonaka, Osaka 560-8531, Japan}
\affiliation{Center for Quantum Information and Quantum Biology, Osaka University, Toyonaka, Osaka 560-0043, Japan}

\begin{abstract}
We propose a linear optical quantum computation scheme using time-frequency degree of freedom.
In this scheme, a qubit is encoded in single-photon frequency combs, and manipulation of the qubits is performed using time-resolving detectors, beam splitters, and optical interleavers.
This scheme does not require active devices such as high-speed switches and electro-optic modulators and is robust against temporal and spectral errors, which are mainly caused by the detectors' finite resolution.
We show that current technologies almost meet the requirements for fault-tolerant quantum computation.
\end{abstract}

\maketitle

\textit{Introduction.---}
Photons and their manipulation using linear optics play an indispensable role in quantum information processing~\cite{Knill2001-eq,Kok2007-iv}.
There has been considerable interest in the choice of the degrees of freedom (d.o.f.) of photons as quantum information carriers~\cite{Mair2001-sc,Humphreys2013-xe,Lukens2014-hb,Brecht2015-zu,Lukens2017-bb}.
The use of time-frequency d.o.f. has several advantages.
First, qubits formed by time-frequency d.o.f. are usually less susceptible to errors because most optical components do not depend on small temporal and spectral differences.
Second, time-frequency d.o.f. is suitable for realizing high-dimensional quantum information processing with qudits because it is a continuous variable. 

There are variations pertaining to encoding using the time-frequency d.o.f.
In time-bin encoding, the temporal peaks of a photon form the computational basis, and its manipulation has been demonstrated by a series of fast switches via spatial or polarization d.o.f.~\cite{Humphreys2013-xe,Takesue2014-nw,Lo2018-zl, Lo2020-qx}.
In frequency-bin encoding, the spectral peaks of a photon form the computational basis, and its manipulation has been demonstrated by a series of electro-optic modulators  and pulse shapers~\cite{Kues2017-un,Lu2018-mg, Lu2019-ig, Lu2018-bh, Lu2020-xg}.
However, the use of many active devices in these approaches is prone to errors and losses and poses challenges in scaling up.
Instead, the manipulation of frequency-bin qubits using time-resolving detectors was recently proposed~\cite{Cui2020-zh}, but the finite resolution of these detectors causes serious errors because frequency-bin qudits are susceptible to temporal shift errors.

In this Letter, we propose a new scheme for linear optical quantum computation (LOQC) using time-frequency d.o.f..
We use encoding in which single-photon frequency combs form the computational basis.
The state in this encoding is called the time-frequency Gottesman-Kitaev-Preskill (TFGKP) state~\cite{Fabre2020-qj,Fabre2022-qo} derived from the analog of
GKP code~\cite{Gottesman2001-gx} for quadrature amplitudes of light~\cite{Weedbrook2012-ib}.
The TFGKP state is robust against time- and frequency-shift errors because it is discretized in both the time and frequency domains.
We show that universal quantum computation can be achieved using TFGKP-state generators, time-resolving detectors, beam splitters (BSs), and optical interleavers (OIs).
Thus, active devices such as high-speed switches and electro-optic modulators are not required.
TFGKP-state generators are efficiently realized using a cavity-enhanced nonlinear optical process~\cite{Reimer2016-ub, Imany2018-ae, Reimer2018-ht, Imany2019-fk, Kues2019-um, Ikuta2019-jb, Maltese2020-et, Yamazaki2022-vv}.
Furthermore, in contrast to the passive scheme that uses frequency-bin encoding~\cite{Cui2020-zh},
quantum computation can be performed robustly despite the detectors' finite resolutions and other temporal and spectral errors.
We estimate the errors occurring in this scheme and show that the experimental requirements for fault-tolerant quantum computation are almost achievable with current state-of-the-art technologies.

\textit{Time-frequency d.o.f. of a photon.---}
We first summarize the expressions and properties of the time-frequency d.o.f. of a photon.
We consider that all probability density functions (PDFs) are localized at the origin.
A complex function $f(\omega)$ is referred to as a probability amplitude function (PAF) when $\abs{f(\omega)}^2$ is a PDF.
We represent the Fourier transformation of a function $f$ by $\hat{f}$ and the pointwise product and convolution of functions $f$ and $g$ by $f\odot g$ and $f*g$, respectively.
We introduce the functions $T_{\omega'}$ and $M_{\tau'}$ as 
\begin{equation}
    T_{\omega'}(f)(\omega)=f(\omega+\omega'), \  M_{\tau'}(f)(\omega)=e^{-i \omega \tau'}f(\omega).
\end{equation}
The annihilation and creation operators of a photon with frequency $\omega$ are represented as $a(\omega)$ and $a^\dag(\omega)$, respectively.
The Fourier transformation of $a(\omega)$ and its adjoint $\hat{a}(\tau)$ and $\hat{a}^\dag(\tau)$ represent the annihilation and creation operators of the photon that arrives at time $\tau$ at a certain point.
Propagation with distance $L$ corresponds to the change in creation operators as $a^\dag(\omega) \rightarrow a^\dag(\omega)e^{-ik(\omega)L}$, where $k$ represents the wave number. 

In practice, photons have finite temporal and spectral widths as wave packets.
A photon wave packet with central frequency $\omega_0$ can be described using PAF $\xi$ as $(\xi * a^\dag)(\omega_0)\ket*{0}$.
Assuming that PAF $\xi$ is sufficiently localized, we can approximate $k$ around $\omega_0$ to the first order as $k(\omega)\simeq k(\omega_0)+k'(\omega-\omega_0)$.
Omitting the constant phase, the propagation of the photon wave packet with distance $L$ corresponds to the transformation $\xi \rightarrow M_{\tau_0} (\xi) $, where $\tau_0=k' L$ is the propagation time.
The state after propagation time $\tau_0$ is 
\begin{equation}\label{wavepacket}
    (M_{\tau_0}(\xi) * a^\dag)(\omega_0)\ket*{0} = e^{i\omega_0\tau_0}(M_{-\omega_0}(\hat{\xi})*\hat{a}^\dag)(\tau_0)\ket*{0}.
\end{equation}
The right side of this equation denotes the temporal photon wave packet centered at $\tau_0$~\footnote[10]{See the Supplemental material for its derivation and details}.

A qudit of time-frequency d.o.f. is affected by unitary and non-unitary errors.
One of the major unitary errors is caused by group velocity dispersion (GVD).
Accounting for the effect of GVD, propagation with distance $L$ corresponds to the transformation of $\xi \rightarrow D\odot M_{\tau_0} (\xi)$, where $D(\omega) = e^{-ik'' L \omega^2/2}$.
Typically, this corresponds to coherent temporal broadening by $\sim \sqrt{8\ln{2} k''L}$.
Coherent spectral broadenings rarely occur except during manipulation.
Probabilistic temporal/spectral shifts can be represented as incoherent temporal/spectral broadening by using a PDF.

Consider the frequency-bin qudit as an example.
It is robust against incoherent spectral broadening because each bin is spectrally isolated unless the bins overlap owing to the broadening.
However, incoherent temporal broadening causes fluctuations in the relative phase between bins.
The error due to this fluctuation is not small even if the broadening is relatively small compared with the inverse of the frequency difference between bins.
Therefore, frequency-bin qudits are susceptible to temporal errors.

\textit{Time-frequency GKP qudit.---}
The GKP and TFGKP qudits were introduced by assuming that the PAFs were Gaussian~\cite{Gottesman2001-gx,Matsuura2020-qm,Fabre2020-qj}.
By contrast, we introduce TFGKP qudits without the Gaussian PAF assumption.
The frequency basis states of the ideal $d$-dimensional TFGKP qudit are defined by the frequency combs formed by a photon with shifted central frequencies.
Using a Dirac comb, that is, the sum of shifted Dirac delta functions $C_{\omega_r}(\omega)=\sum_{n\in\mathbb{Z}}\delta(\omega+n\omega_r)$, they are represented as
\begin{equation}
    \ket*{\overline{j}_f} = (C_{\omega_r}*a^\dag) \qty\big(\frac{j}{d}\omega_r) \ket*{0} = (T_{j\omega_r/d}(C_{\omega_r})*a^\dag) (0) \ket*{0}
    \label{qudit_freq}
\end{equation}
for $j=0,\cdots,d-1$.
Each frequency basis state $\ket*{\overline{j}_f}$ differs from $\ket*{\overline{0}_f}$ by the frequency offset $(j/d)\omega_r$.
The time-basis states of the TFGKP qudit are defined by the discrete Fourier transformation of the frequency-basis states as
\begin{equation}
    \ket*{\overline{j}_t} \equiv \frac{1}{\sqrt{d}}\sum_k e^{-i2\pi j k/d} \ket*{\overline{k}_f} = (C_{\tau_r} * \hat{a}^\dag)\qty\big(\frac{j}{d}\tau_r) \ket*{0}
    \label{qudit_temp}
\end{equation}
for $\tau_r = (2\pi d/\omega_r$ and $j=0,\cdots,d-1$.
$\ket*{\overline{j}_t}$ forms the temporal comb with time period $\tau_r$ and time offset $(j/d)\tau_r$.
Since the Fourier transformation of the Dirac comb is another Dirac comb, this encoding discretizes the states in both the frequency and time domains.

\begin{figure}[tbp]
    \centering
    \includegraphics[scale=0.6, trim=0 0 0 0, clip]{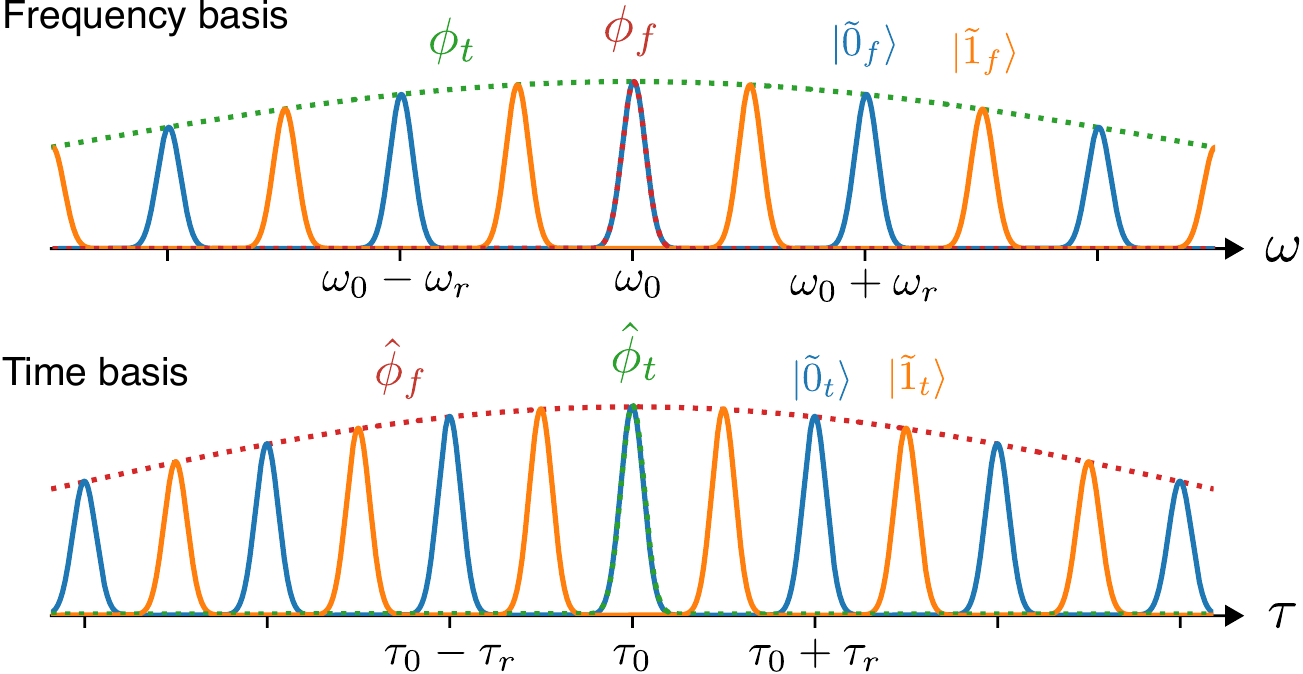} 
    \caption{Probability distributions of a time-frequency Gottesmann--Kitaev--Preskill (TFGKP) qudit in the frequency and time bases for $d=2$. The blue and orange lines in the frequency and time basis represent $\ket*{\tilde{0}_{f/t}}$ and $\ket*{\tilde{1}_{f/t}}$, respectively.}
    \label{example_GKP}
\end{figure}
The states in Eq.~\eqref{qudit_freq} are not normalizable in two respects.
First, each peak consists of a monochromatic mode $a^\dag((n+\frac{j}{d})\omega_r) \ket*{0}$, and second, the summation of the peaks is performed over an infinite range.
To deal with realistic situations, we introduce PAFs $\phi_f$ and $\phi_t$ to represent the spectral broadening of each peak and the envelope of the peaks, respectively.
This corresponds to the replacement $T_{j\omega_r/d}(C_{\omega_r})$ in Eq.~\eqref{qudit_freq} using $\phi_t \odot (T_{j\omega_r/d}(C_{\omega_r})*\phi_f)$.
For central frequency $\omega_0$, the frequency basis states after propagation for time $\tau_0$ are
\begin{equation}
    \ket*{\tilde{j}_f} \propto (\{M_{\tau_0}(\phi_t)\odot[T_{j\omega_r/d}(C_{\omega_r})*\phi_f]\}*a^\dag) (\omega_0) \ket*{0}.
\end{equation}
Then, the time-basis states are~\footnote[10]{See the Supplemental material for its derivation and details.}
\begin{align}
    \ket*{\tilde{j}_t} &\propto \sum_k e^{-i2\pi j k} \ket*{\tilde{k}_f} \notag \\
    &= e^{i\omega_0 \tau_0}(M_{-\omega_0}\{[T_{j\tau_r /d}(C_{\tau_r})\odot \hat{\phi_f})* \hat{\phi_t}\}*\hat{a}^\dag)(\tau_0)\ket*{0}.
    \label{physical_qudit_temp}
\end{align}
We call these normalizable states physical TFGKP states in contrast with the ideal ones.
Figure~\ref{example_GKP} shows an example of physical TFGKP states.
Coherent broadenings of the envelope on the frequency basis are equivalent to coherent compressions of each peak on the time basis and vice versa.

A comb-shaped structure in the frequency domain of light often appears as a series of the transmission peaks of a cavity.
For example, the heralded generation of a TFGKP state is possible using a broadband time-frequency entangled photon pair and a cavity.
When one of the two photons passes through the cavity and is detected by a time-resolving detector, the state of the other photon corresponds to a TFGKP state.
A cavity-enhanced nonlinear optical process~\cite{Reimer2016-ub,Imany2018-ae,Reimer2018-ht,Imany2019-fk,Kues2019-um,Ikuta2019-jb,Maltese2020-et,Yamazaki2022-vv} is applied to the efficient generation of a TFGKP state in this manner~\footnote{Another way to generate a TFGKP state is to use a deterministic broadband single-photon generator and a cavity.
The spectrum of the photons generated inside the cavity corresponds to the transmission spectrum of the cavity.
Therefore, we can generate the TFGKP state in a deterministic manner.
However, the bandwidth of the photon generated by a quantum dot is usually not sufficiently large~\cite{Kuhlmann2015-zs}.}.
By letting the free spectral range (FSR) of the cavity be $\Delta_\text{FSR}$, the generated state corresponds to $\ket*{\tilde{0}_f}$ for $\omega_r=\Delta_\text{FSR}$.
In addition, by setting the frequency period to $\omega_r=d\Delta_\text{FSR}$, the generated state corresponds to $\ket*{\tilde{0}_t}$ for any dimension $d$.
A major incoherent temporal broadening is caused by the finite temporal resolution of the detector in this state-preparation method.

\textit{Optical components and detectors.---}
We introduce our toolbox for the manipulation of TFGKP qudits, which consists of BSs, OIs, and time- and frequency-resolving detectors.
Herein, BSs generally refer to spatial linear optical circuits that are independent of the other d.o.f. of photons, including time-frequency d.o.f..
As we will see below, 50:50 BSs are sufficient for universal quantum computation.

\begin{figure}[tbp]
    \centering
    \includegraphics[scale=1.2, trim=0 0 0 0, clip]{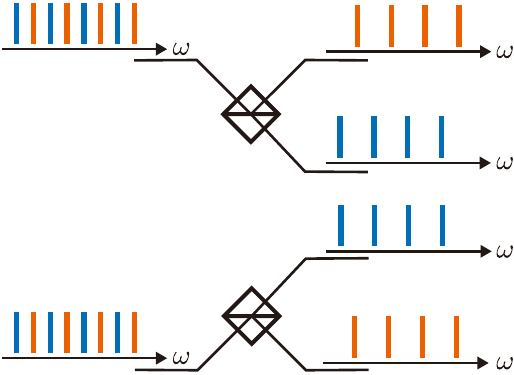}
    \caption{Graphical representation of 2:2 optical interleaver (OI). Orange and blue spectral peaks represent $\ket*{\tilde{0}_f}$ and $\ket*{\tilde{1}_f}$, respectively.} 
    \label{optical_interleaver}
\end{figure}
OI is a spectrally periodic optical filter that spatially combines or separates frequency combs~\cite{Cao2004-iy, Luo2010-kr}.
Herein, we use $d$:$d$ OIs that have $d$ input and output ports~\footnote{A $d$:$d$ OI can be made by connecting $2d$ pieces of commonly used $1$:$d$ OIs.} as shown in Fig.~\ref{optical_interleaver}.
The transformation of the creation operator $a^\dag_j$ in each mode by $d$:$d$ OI is represented as $a^\dag_j\rightarrow \sum_{k=0}^{d-1} I_k \odot a_{{j+k}\pmod{d}}^\dag$, where 
\begin{equation}
    I_j(\omega) = \{g_I \odot [T_{j\omega_r/d}(C_{\omega_r})*f_I]\}(\omega_0-\omega)
\end{equation}
for central frequencies $\omega_0$ and $j=0,\cdots,d-1$.
Functions $g_I$ and $f_I$ represent the transmission coefficients for the envelope and each peak, respectively.
Ideally, an OI routes the frequency basis states of a TFGKP qudit into spatially different $d$ paths~\footnote{A similar functionality can also be implemented by spatially decomposing all the spectral peaks of an input state and recombining a part of them into the same path. However, the method based on an OI that makes effective use of interference would be better.}.

Time- and frequency-resolving detectors are used for photon detection. 
When we detect a photon in state $\ket*{\tilde{i}_t}$ using an ideal time-resolving detector with the infinite resolution, the temporal probability distribution of the photon detection has periodical peaks at $\tau_0 + n(i/d)\tau_r$ for integers $n$.
Even if each peak is blurred by the finite resolution of the detector, we can identify the time-basis states of a TFGKP qudit from which the detection timing is closest unless the finite resolution is as large as the time period.
Similarly, we can identify the frequency basis states of a TFGKP qudit using a frequency-resolving detector with a finite resolution.

A frequency-resolving detector can be substituted by combining a $1$:$d$ OI with $d$ detectors. 
In that case, a discrete result indicating which frequency basis state was detected will be obtained.
By contrast, the use of frequency-resolving detectors, which have been actively studied recently~\cite{Kahl_undated-dj, Cheng2019-na, Young2022-je}, would be advantageous in error analysis~\cite{Fukui2018-ll}.

\begin{figure}[tpb]\label{setups}
    \begin{minipage}[b]{0.55\linewidth}
        \centering
        \subcaption{}
        \includegraphics[scale=1.5, trim=0 0 0 0, clip]{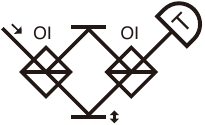}
        \label{magic}
        \subcaption{}
        \includegraphics[scale=1.5, trim=0 0 0 0, clip]{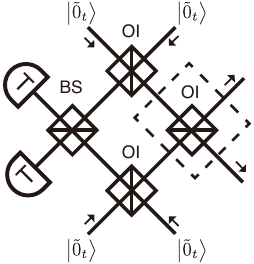}
        \label{Bell_gen}
    \end{minipage}
    \begin{minipage}[b]{0.35\linewidth}
        \centering
        \subcaption{}
        \includegraphics[scale=1.5, trim=0 0 0 0, clip]{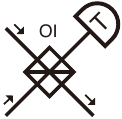} 
        \label{type-1}
        \subcaption{}
        \includegraphics[scale=1.5, trim=0 0 0 0, clip]{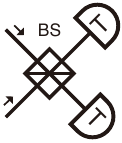}
        \label{type-2'}
        \subcaption{}
        \includegraphics[scale=1.5, trim=0 0 0 0, clip]{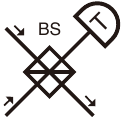}
        \label{type-1'}
    \end{minipage}
    \caption{Concrete setups for quantum operations. All detectors, beam splitters (BSs), and OIs are time-resolving detectors, 50:50 BSs, and $2$:$2$ OIs, respectively. (a) Measurement in the $\cos(\theta/2)X+\sin(\theta/2)Y$ basis. $\theta$ is adjustable by changing the relative lengths between the two arms. (b) Bell-state generation, which succeeds when detectors detect two photons in different states. The OI enclosed by the dotted lines denotes a feed-forward operation required in 1/3 of the success cases. Its total success probability is $3/16$. (c) Type-I fusion gate, which succeeds when a detector detects a photon with a probability $1/2$. (d) Type-II' fusion gate, which succeeds when a detector or detectors detect $\ket*{0}$ and $\ket*{1}$ with a probability of $1/2$. (e) Type-I' fusion gate, which succeeds when a detector detects a photon with a probability of $1/4$.}
    \label{Operations}
\end{figure}
\textit{Universal quantum computation.---}
Let us consider the frequency basis of TFGKP states as the computational basis of a qubit ($d=2$) as $\ket*{0} \simeq \ket*{\tilde{0}_f}$ and $\ket*{1} \simeq \ket*{\tilde{1}_f}$.
The approximation here is due to the use of physical TFGKP states rather than ideal ones.
Then, $\ket{+} = (\ket*{0}+\ket*{1})/\sqrt{2} \simeq \ket*{\tilde{0}_t}$ and $\ket{-} = (\ket*{0}-\ket*{1})/\sqrt{2} \simeq \ket*{\tilde{1}_t}$ hold.
The frequency- and time-resolving measurements correspond to the measurements in the $Z$ and $X$ bases, respectively.
We can realize an arbitrary phase gate by spatially separating each computational basis state by a $1$:$2$ OI, adding a small relative time delay $\Delta \tau $ satisfying $|\Delta \tau| \leq \pi/(2\omega_0) \ll \tau_r$, and then combining them with a $2$:$1$ OI. 
As shown in Fig.~\ref{magic}, the phase gate followed by the measurement in the $X$ basis corresponds to that in the $\cos(\theta/2)X+\sin(\theta/2)Y$ basis.
The measurements in the $Z$ and $X$ bases and the phase gates are readily extendable to qudits with arbitrary dimensions.

By considering measurement-based quantum computation, it is relatively straightforward to demonstrate the quantum computational universality of our toolbox, including TFGKP state generators. In particular, fault-tolerant measurement-based quantum computation can be performed by one-qubit measurements in the $X$, $Z$ and $(X + Y)/\sqrt{2}$ bases on a three-dimensional cluster state~\cite{Raussendorf2007-hp, Raussendorf2006-hf}.
Therefore, the remaining part of achieving universal quantum computation is constructing the cluster state.

For cluster-state generations, we refer to the protocol in polarization encoding using type-I and type-II fusion gates~\cite{Browne2005-ft}; this enables the generation of an arbitrary cluster state from two-qubit cluster states, such as $(\ket*{0+}+\ket*{1-})/\sqrt{2}$.
However, we need to modify it because our toolbox does not include operations corresponding to polarization rotations~\footnote{A type of time-frequency rotation could be realized with active operations such as chirped-pulse up-conversion~\cite{Lavoie2013-wb, Donohue2013-tf}.}.
While we can realize type-I fusion gate as shown in Fig.~\ref{type-1}, we need to introduce type-II' fusion gate shown in Fig.~\ref{type-2'} instead of type-II fusion gate.
Type-II' fusion gate works as well as type-II fusion gate, except for $X$ gate on one of the qubits~\footnote{The effect of this extra $X$ gate can be eliminated by the way the measurement results are interpreted.}.
As shown in Fig.~\ref{Bell_gen}, we also can realize a Bell-state generation setup, similar to the setup in polarization encoding~\cite{Zhang2008-th}.
The generated Bell states differ from the two-qubit cluster state by one Hadamard gate. 
Thus, we additionally introduce type-I' gate shown in Fig.~\ref{type-1'}, to generate a three-qubit cluster state from the two Bell states with a success probability of $1/4$.
By combining these operations, arbitrary cluster state can be generated, allowing for universal quantum computation.

\textit{Error analysis.---}
As the error analysis specific to this scheme, we calculate the errors on the qubits caused by temporal and spectral broadenings.
We call the insufficient separation between states corresponding to the different basis states ``factor I.''
This causes flips in the measurement results and decreases in the success probabilities of the entangling gates.
This is characterized by the total amount of coherent and incoherent broadenings.
By contrast, we call the insufficient overlap between states corresponding to the same basis state ``factor II.''
This degrades the entangling gates and phase gates.
This is characterized by the amount of incoherent broadening relative to that of coherent broadening. 
In our scheme, we use frequency-resolving measurements only for one-qubit measurements; therefore, we do not have to consider factor II on the frequency basis.
On the time basis, there is an optimal amount of coherent temporal broadening owing to a tradeoff between factors I and II.

Herein, we consider only the computational errors. 
We make several assumptions to obtain specific error thresholds. 
We assume that the coherent spectral broadening is characterized by a Lorentzian because it is mainly determined by the transmission spectrum of the cavity used in the state preparation.
This assumption considerably increases the amount of errors compared with the Gaussian assumption.
For simplicity, we ignore the influences of using OIs instead of frequency-resolving detectors and the incoherent spectral broadening~\footnote{This assumption would cause only quantitative differences since there is no need to consider factor II in the frequency basis.}.
Conversely, we assume that coherent and incoherent temporal broadenings are characterized by Gaussian functions~\footnote{On the state preparation using time-frequency entanglement, the shape of coherent temporal broadening can be designed depending on the filter used.}.
The condition subject to which the major error probabilities are less than $0.01$ corresponds to the following conditions~\footnote[10]{See the supplemental material for its derivation and details.},
\begin{equation}\label{errorthereshold}
    \frac{\Delta_{t,i}}{\Delta_{t,c}}\lesssim 0.202,\quad \frac{\Delta_{t,c}}{\tau_r/d} \lesssim 0.476,\quad \frac{\Delta_{f,c}}{\omega_r/d} \lesssim 0.016.
\end{equation}
$\Delta_{t,i}$, $\Delta_{t,c}$, and $\Delta_{f,c}$ are the FWHM values of the incoherent temporal, coherent temporal, and coherent spectral broadenings, respectively.

Let us assume the use of telecom photons around $1.55$~\si{\mu m}, which corresponds to $\omega_0/2\pi \sim 1.9 \times 10^{2}$~\si{THz}. 
The group indices $n_g=ck'$ and GVDs $k''$ are typically $1.5$ and $-2.3 \times 10^4$~\si{fs^2/m} for an optical fiber~\cite{noauthor_undated-nx} and $4.2$ and $-5.6\times 10^6$~\si{fs^2/m} for a silicon-on-insulator waveguide~\cite{Dulkeith2006-bt}, respectively. 
Using these parameters, the lengths corresponding to $1$~ps of time delay and temporal broadening due to GVD are $2.0 \times 10^{-4}$ and $7.8$~\si{m} for the optical fiber and $7.1\times 10^{-2}$ and $3.2 \times 10^{-2}$~\si{m} for the silicon-on-insulator waveguide, respectively.
By contrast, the best value of the time resolution of a detector is $4.3$~\si{ps} for telecom wavelengths~\cite{Korzh2020-yz}.
Thus, we consider only the resolution of the detectors as the major temporal error source for quantum computational applications~\footnote{An implementation of a phase gate induces a time delay $< \pi/2\omega_0 \sim 1.3$~\si{fs}, which is negligible.}.
For $\Delta_{t,i} = 4.3$~\si{ps}, we obtained the required experimental parameters from Eq~\eqref{errorthereshold} as $\Delta_{t,c} \gtrsim 21.5$~ps, $\omega_r/2\pi \lesssim 21$~\si{GHz}, and $\Delta_{f,c}/2\pi \lesssim 0.33/d \lesssim 0.17$~\si{GHz}. 
The first inequality corresponds to that the FWHM of the spectral envelope of a qudit is $ \lesssim 42$~\si{GHz}.
Therefore, the number of frequency bins and the finesse of a state $\ket*{\tilde{0}_t}$ are approximately equal to 2$d$ and 66, respectively.
OIs with $12.5$~\si{GHz} comb spacings are commercially available~\cite{noauthor_undated-rf}.
The generation of a biphoton frequency comb with finesse $\sim 60$~\cite{Ikuta2019-jb, Yamazaki2022-vv} and FSR $= 12.5$~\si{GHz}~\cite{Fujimoto2022-qn} has been demonstrated using nonlinear optical waveguide resonators.
Thus, the current state-of-the-art technologies largely meet the experimental requirements for fault-tolerant quantum computation based on our scheme.

\textit{Conclusion.---}
We proposed a LOQC scheme with TFGKP state generators, time-resolving detectors, BSs, and OIs, and demonstrated the possibility of fault-tolerant quantum computation with currently achievable technologies.
The discretization in both the time and frequency domains owing to TFGKP qubits leads to error robustness against both temporal and spectral errors.
Furthermore, by treating the time and frequency basis asymmetrically, we realized universal quantum computation without active devices.
In addition, this asymmetric structure enabled this scheme to yield good performance, despite the assumption of a Lorentzian coherent spectral broadening.
Additional optimization of TFGKP state generators and OIs for this scheme would relax the requirements of other devices or increase the dimension of the qudit.
Although we show the universality of this scheme for qubits, that is, $d=2$,
its components can be extended to qudits.
Thus, they are a good platform for realizing the recently developed field of LOQC with qudits~\cite{Paesani2021-qn, Zhang2019-dd, Luo2019-hx}.
This scheme has high error robustness and ease of operation due to its use of time-frequency d.o.f. and passive devices.
Therefore, this is a practical approach, especially for quantum computation with integrated photonic circuits~\cite{Dai2013-uv} and quantum communication requiring multiphoton entangled states~\cite{Azuma2015-fr, Ewert2016-ge, Borregaard2020-jk}.

This work was supported by JST Moonshot R\&D JPMJMS2066, JPMJMS226C, MEXT/JSPS KAKENHI JP20H01839, JP20J20261, JP21H04445, and Asahi Glass Foundation.
\bibliography{paperpile.bib}

\begin{thebibliography}{63}%
\makeatletter
\providecommand \@ifxundefined [1]{%
 \@ifx{#1\undefined}
}%
\providecommand \@ifnum [1]{%
 \ifnum #1\expandafter \@firstoftwo
 \else \expandafter \@secondoftwo
 \fi
}%
\providecommand \@ifx [1]{%
 \ifx #1\expandafter \@firstoftwo
 \else \expandafter \@secondoftwo
 \fi
}%
\providecommand \natexlab [1]{#1}%
\providecommand \enquote  [1]{``#1''}%
\providecommand \bibnamefont  [1]{#1}%
\providecommand \bibfnamefont [1]{#1}%
\providecommand \citenamefont [1]{#1}%
\providecommand \href@noop [0]{\@secondoftwo}%
\providecommand \href [0]{\begingroup \@sanitize@url \@href}%
\providecommand \@href[1]{\@@startlink{#1}\@@href}%
\providecommand \@@href[1]{\endgroup#1\@@endlink}%
\providecommand \@sanitize@url [0]{\catcode `\\12\catcode `\$12\catcode
  `\&12\catcode `\#12\catcode `\^12\catcode `\_12\catcode `\%12\relax}%
\providecommand \@@startlink[1]{}%
\providecommand \@@endlink[0]{}%
\providecommand \url  [0]{\begingroup\@sanitize@url \@url }%
\providecommand \@url [1]{\endgroup\@href {#1}{\urlprefix }}%
\providecommand \urlprefix  [0]{URL }%
\providecommand \Eprint [0]{\href }%
\providecommand \doibase [0]{https://doi.org/}%
\providecommand \selectlanguage [0]{\@gobble}%
\providecommand \bibinfo  [0]{\@secondoftwo}%
\providecommand \bibfield  [0]{\@secondoftwo}%
\providecommand \translation [1]{[#1]}%
\providecommand \BibitemOpen [0]{}%
\providecommand \bibitemStop [0]{}%
\providecommand \bibitemNoStop [0]{.\EOS\space}%
\providecommand \EOS [0]{\spacefactor3000\relax}%
\providecommand \BibitemShut  [1]{\csname bibitem#1\endcsname}%
\let\auto@bib@innerbib\@empty
\bibitem [{\citenamefont {Knill}\ \emph {et~al.}(2001)\citenamefont {Knill},
  \citenamefont {Laflamme},\ and\ \citenamefont {Milburn}}]{Knill2001-eq}%
  \BibitemOpen
  \bibfield  {author} {\bibinfo {author} {\bibfnamefont {E.}~\bibnamefont
  {Knill}}, \bibinfo {author} {\bibfnamefont {R.}~\bibnamefont {Laflamme}},\
  and\ \bibinfo {author} {\bibfnamefont {G.~J.}\ \bibnamefont {Milburn}},\
  }\href {https://doi.org/10.1038/35051009} {\bibfield  {journal} {\bibinfo
  {journal} {Nature}\ }\textbf {\bibinfo {volume} {409}},\ \bibinfo {pages}
  {46} (\bibinfo {year} {2001})}\BibitemShut {NoStop}%
\bibitem [{\citenamefont {Kok}\ \emph {et~al.}(2007)\citenamefont {Kok},
  \citenamefont {Munro}, \citenamefont {Nemoto}, \citenamefont {Ralph},
  \citenamefont {Dowling},\ and\ \citenamefont {Milburn}}]{Kok2007-iv}%
  \BibitemOpen
  \bibfield  {author} {\bibinfo {author} {\bibfnamefont {P.}~\bibnamefont
  {Kok}}, \bibinfo {author} {\bibfnamefont {W.~J.}\ \bibnamefont {Munro}},
  \bibinfo {author} {\bibfnamefont {K.}~\bibnamefont {Nemoto}}, \bibinfo
  {author} {\bibfnamefont {T.~C.}\ \bibnamefont {Ralph}}, \bibinfo {author}
  {\bibfnamefont {J.~P.}\ \bibnamefont {Dowling}},\ and\ \bibinfo {author}
  {\bibfnamefont {G.~J.}\ \bibnamefont {Milburn}},\ }\href
  {https://doi.org/10.1103/RevModPhys.79.135} {\bibfield  {journal} {\bibinfo
  {journal} {Rev. Mod. Phys.}\ }\textbf {\bibinfo {volume} {79}},\ \bibinfo
  {pages} {135} (\bibinfo {year} {2007})}\BibitemShut {NoStop}%
\bibitem [{\citenamefont {Mair}\ \emph {et~al.}(2001)\citenamefont {Mair},
  \citenamefont {Vaziri}, \citenamefont {Weihs},\ and\ \citenamefont
  {Zeilinger}}]{Mair2001-sc}%
  \BibitemOpen
  \bibfield  {author} {\bibinfo {author} {\bibfnamefont {A.}~\bibnamefont
  {Mair}}, \bibinfo {author} {\bibfnamefont {A.}~\bibnamefont {Vaziri}},
  \bibinfo {author} {\bibfnamefont {G.}~\bibnamefont {Weihs}},\ and\ \bibinfo
  {author} {\bibfnamefont {A.}~\bibnamefont {Zeilinger}},\ }\href
  {https://doi.org/10.1038/35085529} {\bibfield  {journal} {\bibinfo  {journal}
  {Nature}\ }\textbf {\bibinfo {volume} {412}},\ \bibinfo {pages} {313}
  (\bibinfo {year} {2001})}\BibitemShut {NoStop}%
\bibitem [{\citenamefont {Humphreys}\ \emph {et~al.}(2013)\citenamefont
  {Humphreys}, \citenamefont {Metcalf}, \citenamefont {Spring}, \citenamefont
  {Moore}, \citenamefont {Jin}, \citenamefont {Barbieri}, \citenamefont
  {Kolthammer},\ and\ \citenamefont {Walmsley}}]{Humphreys2013-xe}%
  \BibitemOpen
  \bibfield  {author} {\bibinfo {author} {\bibfnamefont {P.~C.}\ \bibnamefont
  {Humphreys}}, \bibinfo {author} {\bibfnamefont {B.~J.}\ \bibnamefont
  {Metcalf}}, \bibinfo {author} {\bibfnamefont {J.~B.}\ \bibnamefont {Spring}},
  \bibinfo {author} {\bibfnamefont {M.}~\bibnamefont {Moore}}, \bibinfo
  {author} {\bibfnamefont {X.-M.}\ \bibnamefont {Jin}}, \bibinfo {author}
  {\bibfnamefont {M.}~\bibnamefont {Barbieri}}, \bibinfo {author}
  {\bibfnamefont {W.~S.}\ \bibnamefont {Kolthammer}},\ and\ \bibinfo {author}
  {\bibfnamefont {I.~A.}\ \bibnamefont {Walmsley}},\ }\href
  {https://doi.org/10.1103/PhysRevLett.111.150501} {\bibfield  {journal}
  {\bibinfo  {journal} {Phys. Rev. Lett.}\ }\textbf {\bibinfo {volume} {111}},\
  \bibinfo {pages} {150501} (\bibinfo {year} {2013})}\BibitemShut {NoStop}%
\bibitem [{\citenamefont {Lukens}\ \emph {et~al.}(2014)\citenamefont {Lukens},
  \citenamefont {Dezfooliyan}, \citenamefont {Langrock}, \citenamefont {Fejer},
  \citenamefont {Leaird},\ and\ \citenamefont {Weiner}}]{Lukens2014-hb}%
  \BibitemOpen
  \bibfield  {author} {\bibinfo {author} {\bibfnamefont {J.~M.}\ \bibnamefont
  {Lukens}}, \bibinfo {author} {\bibfnamefont {A.}~\bibnamefont {Dezfooliyan}},
  \bibinfo {author} {\bibfnamefont {C.}~\bibnamefont {Langrock}}, \bibinfo
  {author} {\bibfnamefont {M.~M.}\ \bibnamefont {Fejer}}, \bibinfo {author}
  {\bibfnamefont {D.~E.}\ \bibnamefont {Leaird}},\ and\ \bibinfo {author}
  {\bibfnamefont {A.~M.}\ \bibnamefont {Weiner}},\ }\href
  {https://doi.org/10.1103/PhysRevLett.112.133602} {\bibfield  {journal}
  {\bibinfo  {journal} {Phys. Rev. Lett.}\ }\textbf {\bibinfo {volume} {112}},\
  \bibinfo {pages} {133602} (\bibinfo {year} {2014})}\BibitemShut {NoStop}%
\bibitem [{\citenamefont {Brecht}\ \emph {et~al.}(2015)\citenamefont {Brecht},
  \citenamefont {Reddy}, \citenamefont {Silberhorn},\ and\ \citenamefont
  {Raymer}}]{Brecht2015-zu}%
  \BibitemOpen
  \bibfield  {author} {\bibinfo {author} {\bibfnamefont {B.}~\bibnamefont
  {Brecht}}, \bibinfo {author} {\bibfnamefont {D.~V.}\ \bibnamefont {Reddy}},
  \bibinfo {author} {\bibfnamefont {C.}~\bibnamefont {Silberhorn}},\ and\
  \bibinfo {author} {\bibfnamefont {M.~G.}\ \bibnamefont {Raymer}},\ }\href
  {https://doi.org/10.1103/PhysRevX.5.041017} {\bibfield  {journal} {\bibinfo
  {journal} {Phys. Rev. X}\ }\textbf {\bibinfo {volume} {5}},\ \bibinfo {pages}
  {041017} (\bibinfo {year} {2015})}\BibitemShut {NoStop}%
\bibitem [{\citenamefont {Lukens}\ and\ \citenamefont
  {Lougovski}(2017)}]{Lukens2017-bb}%
  \BibitemOpen
  \bibfield  {author} {\bibinfo {author} {\bibfnamefont {J.~M.}\ \bibnamefont
  {Lukens}}\ and\ \bibinfo {author} {\bibfnamefont {P.}~\bibnamefont
  {Lougovski}},\ }\href {https://doi.org/10.1364/OPTICA.4.000008} {\bibfield
  {journal} {\bibinfo  {journal} {Optica, OPTICA}\ }\textbf {\bibinfo {volume}
  {4}},\ \bibinfo {pages} {8} (\bibinfo {year} {2017})}\BibitemShut {NoStop}%
\bibitem [{\citenamefont {Takesue}(2014)}]{Takesue2014-nw}%
  \BibitemOpen
  \bibfield  {author} {\bibinfo {author} {\bibfnamefont {H.}~\bibnamefont
  {Takesue}},\ }\href {https://doi.org/10.1103/PhysRevA.89.062328} {\bibfield
  {journal} {\bibinfo  {journal} {Phys. Rev. A}\ }\textbf {\bibinfo {volume}
  {89}},\ \bibinfo {pages} {062328} (\bibinfo {year} {2014})}\BibitemShut
  {NoStop}%
\bibitem [{\citenamefont {Lo}\ \emph {et~al.}(2018)\citenamefont {Lo},
  \citenamefont {Ikuta}, \citenamefont {Matsuda}, \citenamefont {Honjo},\ and\
  \citenamefont {Takesue}}]{Lo2018-zl}%
  \BibitemOpen
  \bibfield  {author} {\bibinfo {author} {\bibfnamefont {H.-P.}\ \bibnamefont
  {Lo}}, \bibinfo {author} {\bibfnamefont {T.}~\bibnamefont {Ikuta}}, \bibinfo
  {author} {\bibfnamefont {N.}~\bibnamefont {Matsuda}}, \bibinfo {author}
  {\bibfnamefont {T.}~\bibnamefont {Honjo}},\ and\ \bibinfo {author}
  {\bibfnamefont {H.}~\bibnamefont {Takesue}},\ }\href
  {https://doi.org/10.7567/APEX.11.092801} {\bibfield  {journal} {\bibinfo
  {journal} {Appl. Phys. Express}\ }\textbf {\bibinfo {volume} {11}},\ \bibinfo
  {pages} {092801} (\bibinfo {year} {2018})}\BibitemShut {NoStop}%
\bibitem [{\citenamefont {Lo}\ \emph {et~al.}(2020)\citenamefont {Lo},
  \citenamefont {Ikuta}, \citenamefont {Matsuda}, \citenamefont {Honjo},
  \citenamefont {Munro},\ and\ \citenamefont {Takesue}}]{Lo2020-qx}%
  \BibitemOpen
  \bibfield  {author} {\bibinfo {author} {\bibfnamefont {H.-P.}\ \bibnamefont
  {Lo}}, \bibinfo {author} {\bibfnamefont {T.}~\bibnamefont {Ikuta}}, \bibinfo
  {author} {\bibfnamefont {N.}~\bibnamefont {Matsuda}}, \bibinfo {author}
  {\bibfnamefont {T.}~\bibnamefont {Honjo}}, \bibinfo {author} {\bibfnamefont
  {W.~J.}\ \bibnamefont {Munro}},\ and\ \bibinfo {author} {\bibfnamefont
  {H.}~\bibnamefont {Takesue}},\ }\href
  {https://doi.org/10.1103/PhysRevApplied.13.034013} {\bibfield  {journal}
  {\bibinfo  {journal} {Phys. Rev. Appl.}\ }\textbf {\bibinfo {volume} {13}},\
  \bibinfo {pages} {034013} (\bibinfo {year} {2020})}\BibitemShut {NoStop}%
\bibitem [{\citenamefont {Kues}\ \emph {et~al.}(2017)\citenamefont {Kues},
  \citenamefont {Reimer}, \citenamefont {Roztocki}, \citenamefont {Cort{\'e}s},
  \citenamefont {Sciara}, \citenamefont {Wetzel}, \citenamefont {Zhang},
  \citenamefont {Cino}, \citenamefont {Chu}, \citenamefont {Little},
  \citenamefont {Moss}, \citenamefont {Caspani}, \citenamefont {Aza{\~n}a},\
  and\ \citenamefont {Morandotti}}]{Kues2017-un}%
  \BibitemOpen
  \bibfield  {author} {\bibinfo {author} {\bibfnamefont {M.}~\bibnamefont
  {Kues}}, \bibinfo {author} {\bibfnamefont {C.}~\bibnamefont {Reimer}},
  \bibinfo {author} {\bibfnamefont {P.}~\bibnamefont {Roztocki}}, \bibinfo
  {author} {\bibfnamefont {L.~R.}\ \bibnamefont {Cort{\'e}s}}, \bibinfo
  {author} {\bibfnamefont {S.}~\bibnamefont {Sciara}}, \bibinfo {author}
  {\bibfnamefont {B.}~\bibnamefont {Wetzel}}, \bibinfo {author} {\bibfnamefont
  {Y.}~\bibnamefont {Zhang}}, \bibinfo {author} {\bibfnamefont
  {A.}~\bibnamefont {Cino}}, \bibinfo {author} {\bibfnamefont {S.~T.}\
  \bibnamefont {Chu}}, \bibinfo {author} {\bibfnamefont {B.~E.}\ \bibnamefont
  {Little}}, \bibinfo {author} {\bibfnamefont {D.~J.}\ \bibnamefont {Moss}},
  \bibinfo {author} {\bibfnamefont {L.}~\bibnamefont {Caspani}}, \bibinfo
  {author} {\bibfnamefont {J.}~\bibnamefont {Aza{\~n}a}},\ and\ \bibinfo
  {author} {\bibfnamefont {R.}~\bibnamefont {Morandotti}},\ }\href
  {https://doi.org/10.1038/nature22986} {\bibfield  {journal} {\bibinfo
  {journal} {Nature}\ }\textbf {\bibinfo {volume} {546}},\ \bibinfo {pages}
  {622} (\bibinfo {year} {2017})}\BibitemShut {NoStop}%
\bibitem [{\citenamefont {Lu}\ \emph {et~al.}(2018{\natexlab{a}})\citenamefont
  {Lu}, \citenamefont {Lukens}, \citenamefont {Peters}, \citenamefont {Odele},
  \citenamefont {Leaird}, \citenamefont {Weiner},\ and\ \citenamefont
  {Lougovski}}]{Lu2018-mg}%
  \BibitemOpen
  \bibfield  {author} {\bibinfo {author} {\bibfnamefont {H.-H.}\ \bibnamefont
  {Lu}}, \bibinfo {author} {\bibfnamefont {J.~M.}\ \bibnamefont {Lukens}},
  \bibinfo {author} {\bibfnamefont {N.~A.}\ \bibnamefont {Peters}}, \bibinfo
  {author} {\bibfnamefont {O.~D.}\ \bibnamefont {Odele}}, \bibinfo {author}
  {\bibfnamefont {D.~E.}\ \bibnamefont {Leaird}}, \bibinfo {author}
  {\bibfnamefont {A.~M.}\ \bibnamefont {Weiner}},\ and\ \bibinfo {author}
  {\bibfnamefont {P.}~\bibnamefont {Lougovski}},\ }\href
  {https://doi.org/10.1103/PhysRevLett.120.030502} {\bibfield  {journal}
  {\bibinfo  {journal} {Phys. Rev. Lett.}\ }\textbf {\bibinfo {volume} {120}},\
  \bibinfo {pages} {030502} (\bibinfo {year} {2018}{\natexlab{a}})}\BibitemShut
  {NoStop}%
\bibitem [{\citenamefont {Lu}\ \emph {et~al.}(2019)\citenamefont {Lu},
  \citenamefont {Lukens}, \citenamefont {Williams}, \citenamefont {Imany},
  \citenamefont {Peters}, \citenamefont {Weiner},\ and\ \citenamefont
  {Lougovski}}]{Lu2019-ig}%
  \BibitemOpen
  \bibfield  {author} {\bibinfo {author} {\bibfnamefont {H.-H.}\ \bibnamefont
  {Lu}}, \bibinfo {author} {\bibfnamefont {J.~M.}\ \bibnamefont {Lukens}},
  \bibinfo {author} {\bibfnamefont {B.~P.}\ \bibnamefont {Williams}}, \bibinfo
  {author} {\bibfnamefont {P.}~\bibnamefont {Imany}}, \bibinfo {author}
  {\bibfnamefont {N.~A.}\ \bibnamefont {Peters}}, \bibinfo {author}
  {\bibfnamefont {A.~M.}\ \bibnamefont {Weiner}},\ and\ \bibinfo {author}
  {\bibfnamefont {P.}~\bibnamefont {Lougovski}},\ }\href
  {https://doi.org/10.1038/s41534-019-0137-z} {\bibfield  {journal} {\bibinfo
  {journal} {npj Quantum Information}\ }\textbf {\bibinfo {volume} {5}},\
  \bibinfo {pages} {1} (\bibinfo {year} {2019})}\BibitemShut {NoStop}%
\bibitem [{\citenamefont {Lu}\ \emph {et~al.}(2018{\natexlab{b}})\citenamefont
  {Lu}, \citenamefont {Lukens}, \citenamefont {Peters}, \citenamefont
  {Williams}, \citenamefont {Weiner},\ and\ \citenamefont
  {Lougovski}}]{Lu2018-bh}%
  \BibitemOpen
  \bibfield  {author} {\bibinfo {author} {\bibfnamefont {H.-H.}\ \bibnamefont
  {Lu}}, \bibinfo {author} {\bibfnamefont {J.~M.}\ \bibnamefont {Lukens}},
  \bibinfo {author} {\bibfnamefont {N.~A.}\ \bibnamefont {Peters}}, \bibinfo
  {author} {\bibfnamefont {B.~P.}\ \bibnamefont {Williams}}, \bibinfo {author}
  {\bibfnamefont {A.~M.}\ \bibnamefont {Weiner}},\ and\ \bibinfo {author}
  {\bibfnamefont {P.}~\bibnamefont {Lougovski}},\ }\href
  {https://doi.org/10.1364/OPTICA.5.001455} {\bibfield  {journal} {\bibinfo
  {journal} {Optica, OPTICA}\ }\textbf {\bibinfo {volume} {5}},\ \bibinfo
  {pages} {1455} (\bibinfo {year} {2018}{\natexlab{b}})}\BibitemShut {NoStop}%
\bibitem [{\citenamefont {Lu}\ \emph {et~al.}(2020)\citenamefont {Lu},
  \citenamefont {Simmerman}, \citenamefont {Lougovski}, \citenamefont
  {Weiner},\ and\ \citenamefont {Lukens}}]{Lu2020-xg}%
  \BibitemOpen
  \bibfield  {author} {\bibinfo {author} {\bibfnamefont {H.-H.}\ \bibnamefont
  {Lu}}, \bibinfo {author} {\bibfnamefont {E.~M.}\ \bibnamefont {Simmerman}},
  \bibinfo {author} {\bibfnamefont {P.}~\bibnamefont {Lougovski}}, \bibinfo
  {author} {\bibfnamefont {A.~M.}\ \bibnamefont {Weiner}},\ and\ \bibinfo
  {author} {\bibfnamefont {J.~M.}\ \bibnamefont {Lukens}},\ }\href
  {https://doi.org/10.1103/PhysRevLett.125.120503} {\bibfield  {journal}
  {\bibinfo  {journal} {Phys. Rev. Lett.}\ }\textbf {\bibinfo {volume} {125}},\
  \bibinfo {pages} {120503} (\bibinfo {year} {2020})}\BibitemShut {NoStop}%
\bibitem [{\citenamefont {Cui}\ \emph {et~al.}(2020)\citenamefont {Cui},
  \citenamefont {Seshadreesan}, \citenamefont {Guha},\ and\ \citenamefont
  {Fan}}]{Cui2020-zh}%
  \BibitemOpen
  \bibfield  {author} {\bibinfo {author} {\bibfnamefont {C.}~\bibnamefont
  {Cui}}, \bibinfo {author} {\bibfnamefont {K.~P.}\ \bibnamefont
  {Seshadreesan}}, \bibinfo {author} {\bibfnamefont {S.}~\bibnamefont {Guha}},\
  and\ \bibinfo {author} {\bibfnamefont {L.}~\bibnamefont {Fan}},\ }\href
  {https://doi.org/10.1103/PhysRevLett.124.190502} {\bibfield  {journal}
  {\bibinfo  {journal} {Phys. Rev. Lett.}\ }\textbf {\bibinfo {volume} {124}},\
  \bibinfo {pages} {190502} (\bibinfo {year} {2020})}\BibitemShut {NoStop}%
\bibitem [{\citenamefont {Fabre}\ \emph {et~al.}(2020)\citenamefont {Fabre},
  \citenamefont {Maltese}, \citenamefont {Appas}, \citenamefont {Felicetti},
  \citenamefont {Ketterer}, \citenamefont {Keller}, \citenamefont {Coudreau},
  \citenamefont {Baboux}, \citenamefont {Amanti}, \citenamefont {Ducci},\ and\
  \citenamefont {Milman}}]{Fabre2020-qj}%
  \BibitemOpen
  \bibfield  {author} {\bibinfo {author} {\bibfnamefont {N.}~\bibnamefont
  {Fabre}}, \bibinfo {author} {\bibfnamefont {G.}~\bibnamefont {Maltese}},
  \bibinfo {author} {\bibfnamefont {F.}~\bibnamefont {Appas}}, \bibinfo
  {author} {\bibfnamefont {S.}~\bibnamefont {Felicetti}}, \bibinfo {author}
  {\bibfnamefont {A.}~\bibnamefont {Ketterer}}, \bibinfo {author}
  {\bibfnamefont {A.}~\bibnamefont {Keller}}, \bibinfo {author} {\bibfnamefont
  {T.}~\bibnamefont {Coudreau}}, \bibinfo {author} {\bibfnamefont
  {F.}~\bibnamefont {Baboux}}, \bibinfo {author} {\bibfnamefont {M.~I.}\
  \bibnamefont {Amanti}}, \bibinfo {author} {\bibfnamefont {S.}~\bibnamefont
  {Ducci}},\ and\ \bibinfo {author} {\bibfnamefont {P.}~\bibnamefont
  {Milman}},\ }\href {https://doi.org/10.1103/PhysRevA.102.012607} {\bibfield
  {journal} {\bibinfo  {journal} {Phys. Rev. A}\ }\textbf {\bibinfo {volume}
  {102}},\ \bibinfo {pages} {012607} (\bibinfo {year} {2020})}\BibitemShut
  {NoStop}%
\bibitem [{\citenamefont {Fabre}\ \emph {et~al.}(2022)\citenamefont {Fabre},
  \citenamefont {Keller},\ and\ \citenamefont {Milman}}]{Fabre2022-qo}%
  \BibitemOpen
  \bibfield  {author} {\bibinfo {author} {\bibfnamefont {N.}~\bibnamefont
  {Fabre}}, \bibinfo {author} {\bibfnamefont {A.}~\bibnamefont {Keller}},\ and\
  \bibinfo {author} {\bibfnamefont {P.}~\bibnamefont {Milman}},\ }\href
  {https://doi.org/10.1103/PhysRevA.105.052429} {\bibfield  {journal} {\bibinfo
   {journal} {Phys. Rev. A}\ }\textbf {\bibinfo {volume} {105}},\ \bibinfo
  {pages} {052429} (\bibinfo {year} {2022})}\BibitemShut {NoStop}%
\bibitem [{\citenamefont {Gottesman}\ \emph {et~al.}(2001)\citenamefont
  {Gottesman}, \citenamefont {Kitaev},\ and\ \citenamefont
  {Preskill}}]{Gottesman2001-gx}%
  \BibitemOpen
  \bibfield  {author} {\bibinfo {author} {\bibfnamefont {D.}~\bibnamefont
  {Gottesman}}, \bibinfo {author} {\bibfnamefont {A.}~\bibnamefont {Kitaev}},\
  and\ \bibinfo {author} {\bibfnamefont {J.}~\bibnamefont {Preskill}},\ }\href
  {https://doi.org/10.1103/PhysRevA.64.012310} {\bibfield  {journal} {\bibinfo
  {journal} {Phys. Rev. A}\ }\textbf {\bibinfo {volume} {64}},\ \bibinfo
  {pages} {012310} (\bibinfo {year} {2001})}\BibitemShut {NoStop}%
\bibitem [{\citenamefont {Weedbrook}\ \emph {et~al.}(2012)\citenamefont
  {Weedbrook}, \citenamefont {Pirandola}, \citenamefont
  {Garc{\'\i}a-Patr{\'o}n}, \citenamefont {Cerf}, \citenamefont {Ralph},
  \citenamefont {Shapiro},\ and\ \citenamefont {Lloyd}}]{Weedbrook2012-ib}%
  \BibitemOpen
  \bibfield  {author} {\bibinfo {author} {\bibfnamefont {C.}~\bibnamefont
  {Weedbrook}}, \bibinfo {author} {\bibfnamefont {S.}~\bibnamefont
  {Pirandola}}, \bibinfo {author} {\bibfnamefont {R.}~\bibnamefont
  {Garc{\'\i}a-Patr{\'o}n}}, \bibinfo {author} {\bibfnamefont {N.~J.}\
  \bibnamefont {Cerf}}, \bibinfo {author} {\bibfnamefont {T.~C.}\ \bibnamefont
  {Ralph}}, \bibinfo {author} {\bibfnamefont {J.~H.}\ \bibnamefont {Shapiro}},\
  and\ \bibinfo {author} {\bibfnamefont {S.}~\bibnamefont {Lloyd}},\ }\href
  {https://doi.org/10.1103/RevModPhys.84.621} {\bibfield  {journal} {\bibinfo
  {journal} {Rev. Mod. Phys.}\ }\textbf {\bibinfo {volume} {84}},\ \bibinfo
  {pages} {621} (\bibinfo {year} {2012})}\BibitemShut {NoStop}%
\bibitem [{\citenamefont {Reimer}\ \emph {et~al.}(2016)\citenamefont {Reimer},
  \citenamefont {Kues}, \citenamefont {Roztocki}, \citenamefont {Wetzel},
  \citenamefont {Grazioso}, \citenamefont {Little}, \citenamefont {Chu},
  \citenamefont {Johnston}, \citenamefont {Bromberg}, \citenamefont {Caspani},
  \citenamefont {Moss},\ and\ \citenamefont {Morandotti}}]{Reimer2016-ub}%
  \BibitemOpen
  \bibfield  {author} {\bibinfo {author} {\bibfnamefont {C.}~\bibnamefont
  {Reimer}}, \bibinfo {author} {\bibfnamefont {M.}~\bibnamefont {Kues}},
  \bibinfo {author} {\bibfnamefont {P.}~\bibnamefont {Roztocki}}, \bibinfo
  {author} {\bibfnamefont {B.}~\bibnamefont {Wetzel}}, \bibinfo {author}
  {\bibfnamefont {F.}~\bibnamefont {Grazioso}}, \bibinfo {author}
  {\bibfnamefont {B.~E.}\ \bibnamefont {Little}}, \bibinfo {author}
  {\bibfnamefont {S.~T.}\ \bibnamefont {Chu}}, \bibinfo {author} {\bibfnamefont
  {T.}~\bibnamefont {Johnston}}, \bibinfo {author} {\bibfnamefont
  {Y.}~\bibnamefont {Bromberg}}, \bibinfo {author} {\bibfnamefont
  {L.}~\bibnamefont {Caspani}}, \bibinfo {author} {\bibfnamefont {D.~J.}\
  \bibnamefont {Moss}},\ and\ \bibinfo {author} {\bibfnamefont
  {R.}~\bibnamefont {Morandotti}},\ }\href
  {https://doi.org/10.1126/science.aad8532} {\bibfield  {journal} {\bibinfo
  {journal} {Science}\ }\textbf {\bibinfo {volume} {351}},\ \bibinfo {pages}
  {1176} (\bibinfo {year} {2016})}\BibitemShut {NoStop}%
\bibitem [{\citenamefont {Imany}\ \emph {et~al.}(2018)\citenamefont {Imany},
  \citenamefont {Jaramillo-Villegas}, \citenamefont {Odele}, \citenamefont
  {Han}, \citenamefont {Leaird}, \citenamefont {Lukens}, \citenamefont
  {Lougovski}, \citenamefont {Qi},\ and\ \citenamefont
  {Weiner}}]{Imany2018-ae}%
  \BibitemOpen
  \bibfield  {author} {\bibinfo {author} {\bibfnamefont {P.}~\bibnamefont
  {Imany}}, \bibinfo {author} {\bibfnamefont {J.~A.}\ \bibnamefont
  {Jaramillo-Villegas}}, \bibinfo {author} {\bibfnamefont {O.~D.}\ \bibnamefont
  {Odele}}, \bibinfo {author} {\bibfnamefont {K.}~\bibnamefont {Han}}, \bibinfo
  {author} {\bibfnamefont {D.~E.}\ \bibnamefont {Leaird}}, \bibinfo {author}
  {\bibfnamefont {J.~M.}\ \bibnamefont {Lukens}}, \bibinfo {author}
  {\bibfnamefont {P.}~\bibnamefont {Lougovski}}, \bibinfo {author}
  {\bibfnamefont {M.}~\bibnamefont {Qi}},\ and\ \bibinfo {author}
  {\bibfnamefont {A.~M.}\ \bibnamefont {Weiner}},\ }\href
  {https://doi.org/10.1364/OE.26.001825} {\bibfield  {journal} {\bibinfo
  {journal} {Opt. Express}\ }\textbf {\bibinfo {volume} {26}},\ \bibinfo
  {pages} {1825} (\bibinfo {year} {2018})}\BibitemShut {NoStop}%
\bibitem [{\citenamefont {Reimer}\ \emph {et~al.}(2018)\citenamefont {Reimer},
  \citenamefont {Sciara}, \citenamefont {Roztocki}, \citenamefont {Islam},
  \citenamefont {Romero~Cort{\'e}s}, \citenamefont {Zhang}, \citenamefont
  {Fischer}, \citenamefont {Loranger}, \citenamefont {Kashyap}, \citenamefont
  {Cino}, \citenamefont {Chu}, \citenamefont {Little}, \citenamefont {Moss},
  \citenamefont {Caspani}, \citenamefont {Munro}, \citenamefont {Aza{\~n}a},
  \citenamefont {Kues},\ and\ \citenamefont {Morandotti}}]{Reimer2018-ht}%
  \BibitemOpen
  \bibfield  {author} {\bibinfo {author} {\bibfnamefont {C.}~\bibnamefont
  {Reimer}}, \bibinfo {author} {\bibfnamefont {S.}~\bibnamefont {Sciara}},
  \bibinfo {author} {\bibfnamefont {P.}~\bibnamefont {Roztocki}}, \bibinfo
  {author} {\bibfnamefont {M.}~\bibnamefont {Islam}}, \bibinfo {author}
  {\bibfnamefont {L.}~\bibnamefont {Romero~Cort{\'e}s}}, \bibinfo {author}
  {\bibfnamefont {Y.}~\bibnamefont {Zhang}}, \bibinfo {author} {\bibfnamefont
  {B.}~\bibnamefont {Fischer}}, \bibinfo {author} {\bibfnamefont
  {S.}~\bibnamefont {Loranger}}, \bibinfo {author} {\bibfnamefont
  {R.}~\bibnamefont {Kashyap}}, \bibinfo {author} {\bibfnamefont
  {A.}~\bibnamefont {Cino}}, \bibinfo {author} {\bibfnamefont {S.~T.}\
  \bibnamefont {Chu}}, \bibinfo {author} {\bibfnamefont {B.~E.}\ \bibnamefont
  {Little}}, \bibinfo {author} {\bibfnamefont {D.~J.}\ \bibnamefont {Moss}},
  \bibinfo {author} {\bibfnamefont {L.}~\bibnamefont {Caspani}}, \bibinfo
  {author} {\bibfnamefont {W.~J.}\ \bibnamefont {Munro}}, \bibinfo {author}
  {\bibfnamefont {J.}~\bibnamefont {Aza{\~n}a}}, \bibinfo {author}
  {\bibfnamefont {M.}~\bibnamefont {Kues}},\ and\ \bibinfo {author}
  {\bibfnamefont {R.}~\bibnamefont {Morandotti}},\ }\href
  {https://doi.org/10.1038/s41567-018-0347-x} {\bibfield  {journal} {\bibinfo
  {journal} {Nat. Phys.}\ }\textbf {\bibinfo {volume} {15}},\ \bibinfo {pages}
  {148} (\bibinfo {year} {2018})}\BibitemShut {NoStop}%
\bibitem [{\citenamefont {Imany}\ \emph {et~al.}(2019)\citenamefont {Imany},
  \citenamefont {Jaramillo-Villegas}, \citenamefont {Alshaykh}, \citenamefont
  {Lukens}, \citenamefont {Odele}, \citenamefont {Moore}, \citenamefont
  {Leaird}, \citenamefont {Qi},\ and\ \citenamefont {Weiner}}]{Imany2019-fk}%
  \BibitemOpen
  \bibfield  {author} {\bibinfo {author} {\bibfnamefont {P.}~\bibnamefont
  {Imany}}, \bibinfo {author} {\bibfnamefont {J.~A.}\ \bibnamefont
  {Jaramillo-Villegas}}, \bibinfo {author} {\bibfnamefont {M.~S.}\ \bibnamefont
  {Alshaykh}}, \bibinfo {author} {\bibfnamefont {J.~M.}\ \bibnamefont
  {Lukens}}, \bibinfo {author} {\bibfnamefont {O.~D.}\ \bibnamefont {Odele}},
  \bibinfo {author} {\bibfnamefont {A.~J.}\ \bibnamefont {Moore}}, \bibinfo
  {author} {\bibfnamefont {D.~E.}\ \bibnamefont {Leaird}}, \bibinfo {author}
  {\bibfnamefont {M.}~\bibnamefont {Qi}},\ and\ \bibinfo {author}
  {\bibfnamefont {A.~M.}\ \bibnamefont {Weiner}},\ }\href
  {https://doi.org/10.1038/s41534-019-0173-8} {\bibfield  {journal} {\bibinfo
  {journal} {npj Quantum Information}\ }\textbf {\bibinfo {volume} {5}},\
  \bibinfo {pages} {1} (\bibinfo {year} {2019})}\BibitemShut {NoStop}%
\bibitem [{\citenamefont {Kues}\ \emph {et~al.}(2019)\citenamefont {Kues},
  \citenamefont {Reimer}, \citenamefont {Lukens}, \citenamefont {Munro},
  \citenamefont {Weiner}, \citenamefont {Moss},\ and\ \citenamefont
  {Morandotti}}]{Kues2019-um}%
  \BibitemOpen
  \bibfield  {author} {\bibinfo {author} {\bibfnamefont {M.}~\bibnamefont
  {Kues}}, \bibinfo {author} {\bibfnamefont {C.}~\bibnamefont {Reimer}},
  \bibinfo {author} {\bibfnamefont {J.~M.}\ \bibnamefont {Lukens}}, \bibinfo
  {author} {\bibfnamefont {W.~J.}\ \bibnamefont {Munro}}, \bibinfo {author}
  {\bibfnamefont {A.~M.}\ \bibnamefont {Weiner}}, \bibinfo {author}
  {\bibfnamefont {D.~J.}\ \bibnamefont {Moss}},\ and\ \bibinfo {author}
  {\bibfnamefont {R.}~\bibnamefont {Morandotti}},\ }\href
  {https://doi.org/10.1038/s41566-019-0363-0} {\bibfield  {journal} {\bibinfo
  {journal} {Nat. Photonics}\ }\textbf {\bibinfo {volume} {13}},\ \bibinfo
  {pages} {170} (\bibinfo {year} {2019})}\BibitemShut {NoStop}%
\bibitem [{\citenamefont {Ikuta}\ \emph {et~al.}(2019)\citenamefont {Ikuta},
  \citenamefont {Tani}, \citenamefont {Ishizaki}, \citenamefont {Miki},
  \citenamefont {Yabuno}, \citenamefont {Terai}, \citenamefont {Imoto},\ and\
  \citenamefont {Yamamoto}}]{Ikuta2019-jb}%
  \BibitemOpen
  \bibfield  {author} {\bibinfo {author} {\bibfnamefont {R.}~\bibnamefont
  {Ikuta}}, \bibinfo {author} {\bibfnamefont {R.}~\bibnamefont {Tani}},
  \bibinfo {author} {\bibfnamefont {M.}~\bibnamefont {Ishizaki}}, \bibinfo
  {author} {\bibfnamefont {S.}~\bibnamefont {Miki}}, \bibinfo {author}
  {\bibfnamefont {M.}~\bibnamefont {Yabuno}}, \bibinfo {author} {\bibfnamefont
  {H.}~\bibnamefont {Terai}}, \bibinfo {author} {\bibfnamefont
  {N.}~\bibnamefont {Imoto}},\ and\ \bibinfo {author} {\bibfnamefont
  {T.}~\bibnamefont {Yamamoto}},\ }\href
  {https://doi.org/10.1103/PhysRevLett.123.193603} {\bibfield  {journal}
  {\bibinfo  {journal} {Phys. Rev. Lett.}\ }\textbf {\bibinfo {volume} {123}},\
  \bibinfo {pages} {193603} (\bibinfo {year} {2019})}\BibitemShut {NoStop}%
\bibitem [{\citenamefont {Maltese}\ \emph {et~al.}(2020)\citenamefont
  {Maltese}, \citenamefont {Amanti}, \citenamefont {Appas}, \citenamefont
  {Sinnl}, \citenamefont {Lema{\^\i}tre}, \citenamefont {Milman}, \citenamefont
  {Baboux},\ and\ \citenamefont {Ducci}}]{Maltese2020-et}%
  \BibitemOpen
  \bibfield  {author} {\bibinfo {author} {\bibfnamefont {G.}~\bibnamefont
  {Maltese}}, \bibinfo {author} {\bibfnamefont {M.~I.}\ \bibnamefont {Amanti}},
  \bibinfo {author} {\bibfnamefont {F.}~\bibnamefont {Appas}}, \bibinfo
  {author} {\bibfnamefont {G.}~\bibnamefont {Sinnl}}, \bibinfo {author}
  {\bibfnamefont {A.}~\bibnamefont {Lema{\^\i}tre}}, \bibinfo {author}
  {\bibfnamefont {P.}~\bibnamefont {Milman}}, \bibinfo {author} {\bibfnamefont
  {F.}~\bibnamefont {Baboux}},\ and\ \bibinfo {author} {\bibfnamefont
  {S.}~\bibnamefont {Ducci}},\ }\href
  {https://doi.org/10.1038/s41534-019-0237-9} {\bibfield  {journal} {\bibinfo
  {journal} {npj Quantum Information}\ }\textbf {\bibinfo {volume} {6}},\
  \bibinfo {pages} {1} (\bibinfo {year} {2020})}\BibitemShut {NoStop}%
\bibitem [{\citenamefont {Yamazaki}\ \emph {et~al.}(2022)\citenamefont
  {Yamazaki}, \citenamefont {Ikuta}, \citenamefont {Kobayashi}, \citenamefont
  {Miki}, \citenamefont {China}, \citenamefont {Terai}, \citenamefont {Imoto},\
  and\ \citenamefont {Yamamoto}}]{Yamazaki2022-vv}%
  \BibitemOpen
  \bibfield  {author} {\bibinfo {author} {\bibfnamefont {T.}~\bibnamefont
  {Yamazaki}}, \bibinfo {author} {\bibfnamefont {R.}~\bibnamefont {Ikuta}},
  \bibinfo {author} {\bibfnamefont {T.}~\bibnamefont {Kobayashi}}, \bibinfo
  {author} {\bibfnamefont {S.}~\bibnamefont {Miki}}, \bibinfo {author}
  {\bibfnamefont {F.}~\bibnamefont {China}}, \bibinfo {author} {\bibfnamefont
  {H.}~\bibnamefont {Terai}}, \bibinfo {author} {\bibfnamefont
  {N.}~\bibnamefont {Imoto}},\ and\ \bibinfo {author} {\bibfnamefont
  {T.}~\bibnamefont {Yamamoto}},\ }\href
  {https://doi.org/10.1038/s41598-022-12691-7} {\bibfield  {journal} {\bibinfo
  {journal} {Sci. Rep.}\ }\textbf {\bibinfo {volume} {12}},\ \bibinfo {pages}
  {8964} (\bibinfo {year} {2022})}\BibitemShut {NoStop}%
\bibitem [{Note10()}]{Note10}%
  \BibitemOpen
  \bibinfo {note} {See the Supplemental material for its derivation and
  details}\BibitemShut {NoStop}%
\bibitem [{\citenamefont {Matsuura}\ \emph {et~al.}(2020)\citenamefont
  {Matsuura}, \citenamefont {Yamasaki},\ and\ \citenamefont
  {Koashi}}]{Matsuura2020-qm}%
  \BibitemOpen
  \bibfield  {author} {\bibinfo {author} {\bibfnamefont {T.}~\bibnamefont
  {Matsuura}}, \bibinfo {author} {\bibfnamefont {H.}~\bibnamefont {Yamasaki}},\
  and\ \bibinfo {author} {\bibfnamefont {M.}~\bibnamefont {Koashi}},\ }\href
  {https://doi.org/10.1103/PhysRevA.102.032408} {\bibfield  {journal} {\bibinfo
   {journal} {Phys. Rev. A}\ }\textbf {\bibinfo {volume} {102}},\ \bibinfo
  {pages} {032408} (\bibinfo {year} {2020})}\BibitemShut {NoStop}%
\bibitem [{Note1()}]{Note1}%
  \BibitemOpen
  \bibinfo {note} {Another way to generate a TFGKP state is to use a
  deterministic broadband single-photon generator and a cavity. The spectrum of
  the photons generated inside the cavity corresponds to the transmission
  spectrum of the cavity. Therefore, we can generate the TFGKP state in a
  deterministic manner. However, the bandwidth of the photon generated by a
  quantum dot is usually not sufficiently large~\cite
  {Kuhlmann2015-zs}.}\BibitemShut {Stop}%
\bibitem [{\citenamefont {Cao}\ \emph {et~al.}(2004)\citenamefont {Cao},
  \citenamefont {Chen}, \citenamefont {Damask}, \citenamefont {Doerr},
  \citenamefont {Guiziou}, \citenamefont {Harvey}, \citenamefont {Hibino},
  \citenamefont {Li}, \citenamefont {Suzuki}, \citenamefont {Wu},\ and\
  \citenamefont {Xie}}]{Cao2004-iy}%
  \BibitemOpen
  \bibfield  {author} {\bibinfo {author} {\bibfnamefont {S.}~\bibnamefont
  {Cao}}, \bibinfo {author} {\bibfnamefont {J.}~\bibnamefont {Chen}}, \bibinfo
  {author} {\bibfnamefont {J.~N.}\ \bibnamefont {Damask}}, \bibinfo {author}
  {\bibfnamefont {C.~R.}\ \bibnamefont {Doerr}}, \bibinfo {author}
  {\bibfnamefont {L.}~\bibnamefont {Guiziou}}, \bibinfo {author} {\bibfnamefont
  {G.}~\bibnamefont {Harvey}}, \bibinfo {author} {\bibfnamefont
  {Y.}~\bibnamefont {Hibino}}, \bibinfo {author} {\bibfnamefont
  {H.}~\bibnamefont {Li}}, \bibinfo {author} {\bibfnamefont {S.}~\bibnamefont
  {Suzuki}}, \bibinfo {author} {\bibfnamefont {K.-Y.}\ \bibnamefont {Wu}},\
  and\ \bibinfo {author} {\bibfnamefont {P.}~\bibnamefont {Xie}},\ }\href
  {https://doi.org/10.1109/JLT.2003.822832} {\bibfield  {journal} {\bibinfo
  {journal} {J. Lightwave Technol.}\ }\textbf {\bibinfo {volume} {22}},\
  \bibinfo {pages} {281} (\bibinfo {year} {2004})}\BibitemShut {NoStop}%
\bibitem [{\citenamefont {Luo}\ \emph {et~al.}(2010)\citenamefont {Luo},
  \citenamefont {Ibrahim}, \citenamefont {Nitkowski}, \citenamefont {Ding},
  \citenamefont {Poitras}, \citenamefont {Ben~Yoo},\ and\ \citenamefont
  {Lipson}}]{Luo2010-kr}%
  \BibitemOpen
  \bibfield  {author} {\bibinfo {author} {\bibfnamefont {L.-W.}\ \bibnamefont
  {Luo}}, \bibinfo {author} {\bibfnamefont {S.}~\bibnamefont {Ibrahim}},
  \bibinfo {author} {\bibfnamefont {A.}~\bibnamefont {Nitkowski}}, \bibinfo
  {author} {\bibfnamefont {Z.}~\bibnamefont {Ding}}, \bibinfo {author}
  {\bibfnamefont {C.~B.}\ \bibnamefont {Poitras}}, \bibinfo {author}
  {\bibfnamefont {S.~J.}\ \bibnamefont {Ben~Yoo}},\ and\ \bibinfo {author}
  {\bibfnamefont {M.}~\bibnamefont {Lipson}},\ }\href
  {https://doi.org/10.1364/OE.18.023079} {\bibfield  {journal} {\bibinfo
  {journal} {Opt. Express}\ }\textbf {\bibinfo {volume} {18}},\ \bibinfo
  {pages} {23079} (\bibinfo {year} {2010})}\BibitemShut {NoStop}%
\bibitem [{Note2()}]{Note2}%
  \BibitemOpen
  \bibinfo {note} {A $d$:$d$ OI can be made by connecting $2d$ pieces of
  commonly used $1$:$d$ OIs.}\BibitemShut {Stop}%
\bibitem [{Note3()}]{Note3}%
  \BibitemOpen
  \bibinfo {note} {A similar functionality can also be implemented by spatially
  decomposing all the spectral peaks of an input state and recombining a part
  of them into the same path. However, the method based on an OI that makes
  effective use of interference would be better.}\BibitemShut {Stop}%
\bibitem [{\citenamefont {Kahl}\ \emph {et~al.}()\citenamefont {Kahl},
  \citenamefont {Ferrari}, \citenamefont {Kovalyuk}, \citenamefont {Vetter},
  \citenamefont {Lewes-Malandrakis}, \citenamefont {Nebel}, \citenamefont
  {Korneev}, \citenamefont {Goltsman},\ and\ \citenamefont
  {Pernice}}]{Kahl_undated-dj}%
  \BibitemOpen
  \bibfield  {author} {\bibinfo {author} {\bibfnamefont {O.}~\bibnamefont
  {Kahl}}, \bibinfo {author} {\bibfnamefont {S.}~\bibnamefont {Ferrari}},
  \bibinfo {author} {\bibfnamefont {V.}~\bibnamefont {Kovalyuk}}, \bibinfo
  {author} {\bibfnamefont {A.}~\bibnamefont {Vetter}}, \bibinfo {author}
  {\bibfnamefont {G.}~\bibnamefont {Lewes-Malandrakis}}, \bibinfo {author}
  {\bibfnamefont {C.}~\bibnamefont {Nebel}}, \bibinfo {author} {\bibfnamefont
  {A.}~\bibnamefont {Korneev}}, \bibinfo {author} {\bibfnamefont
  {G.}~\bibnamefont {Goltsman}},\ and\ \bibinfo {author} {\bibfnamefont
  {W.}~\bibnamefont {Pernice}},\ }\bibfield  {journal} {\bibinfo  {journal}
  {Optica, OPTICA}\ }\href {https://doi.org/10.1364/OPTICA.4.000557}
  {10.1364/OPTICA.4.000557}\BibitemShut {NoStop}%
\bibitem [{\citenamefont {Cheng}\ \emph {et~al.}(2019)\citenamefont {Cheng},
  \citenamefont {Zou}, \citenamefont {Guo}, \citenamefont {Wang}, \citenamefont
  {Han},\ and\ \citenamefont {Tang}}]{Cheng2019-na}%
  \BibitemOpen
  \bibfield  {author} {\bibinfo {author} {\bibfnamefont {R.}~\bibnamefont
  {Cheng}}, \bibinfo {author} {\bibfnamefont {C.-L.}\ \bibnamefont {Zou}},
  \bibinfo {author} {\bibfnamefont {X.}~\bibnamefont {Guo}}, \bibinfo {author}
  {\bibfnamefont {S.}~\bibnamefont {Wang}}, \bibinfo {author} {\bibfnamefont
  {X.}~\bibnamefont {Han}},\ and\ \bibinfo {author} {\bibfnamefont {H.~X.}\
  \bibnamefont {Tang}},\ }\href {https://doi.org/10.1038/s41467-019-12149-x}
  {\bibfield  {journal} {\bibinfo  {journal} {Nat. Commun.}\ }\textbf {\bibinfo
  {volume} {10}},\ \bibinfo {pages} {4104} (\bibinfo {year}
  {2019})}\BibitemShut {NoStop}%
\bibitem [{\citenamefont {Young}\ \emph {et~al.}(2022)\citenamefont {Young},
  \citenamefont {Sarovar},\ and\ \citenamefont {L{\'e}onard}}]{Young2022-je}%
  \BibitemOpen
  \bibfield  {author} {\bibinfo {author} {\bibfnamefont {S.~M.}\ \bibnamefont
  {Young}}, \bibinfo {author} {\bibfnamefont {M.}~\bibnamefont {Sarovar}},\
  and\ \bibinfo {author} {\bibfnamefont {F.}~\bibnamefont {L{\'e}onard}},\
  }\href@noop {} {\  (\bibinfo {year} {2022})},\ \Eprint
  {https://arxiv.org/abs/2205.05817} {arXiv:2205.05817 [quant-ph]} \BibitemShut
  {NoStop}%
\bibitem [{\citenamefont {Fukui}\ \emph {et~al.}(2018)\citenamefont {Fukui},
  \citenamefont {Tomita}, \citenamefont {Okamoto},\ and\ \citenamefont
  {Fujii}}]{Fukui2018-ll}%
  \BibitemOpen
  \bibfield  {author} {\bibinfo {author} {\bibfnamefont {K.}~\bibnamefont
  {Fukui}}, \bibinfo {author} {\bibfnamefont {A.}~\bibnamefont {Tomita}},
  \bibinfo {author} {\bibfnamefont {A.}~\bibnamefont {Okamoto}},\ and\ \bibinfo
  {author} {\bibfnamefont {K.}~\bibnamefont {Fujii}},\ }\href
  {https://doi.org/10.1103/PhysRevX.8.021054} {\bibfield  {journal} {\bibinfo
  {journal} {Phys. Rev. X}\ }\textbf {\bibinfo {volume} {8}},\ \bibinfo {pages}
  {021054} (\bibinfo {year} {2018})}\BibitemShut {NoStop}%
\bibitem [{\citenamefont {Raussendorf}\ and\ \citenamefont
  {Harrington}(2007)}]{Raussendorf2007-hp}%
  \BibitemOpen
  \bibfield  {author} {\bibinfo {author} {\bibfnamefont {R.}~\bibnamefont
  {Raussendorf}}\ and\ \bibinfo {author} {\bibfnamefont {J.}~\bibnamefont
  {Harrington}},\ }\href {https://doi.org/10.1103/PhysRevLett.98.190504}
  {\bibfield  {journal} {\bibinfo  {journal} {Phys. Rev. Lett.}\ }\textbf
  {\bibinfo {volume} {98}},\ \bibinfo {pages} {190504} (\bibinfo {year}
  {2007})}\BibitemShut {NoStop}%
\bibitem [{\citenamefont {Raussendorf}\ \emph {et~al.}(2006)\citenamefont
  {Raussendorf}, \citenamefont {Harrington},\ and\ \citenamefont
  {Goyal}}]{Raussendorf2006-hf}%
  \BibitemOpen
  \bibfield  {author} {\bibinfo {author} {\bibfnamefont {R.}~\bibnamefont
  {Raussendorf}}, \bibinfo {author} {\bibfnamefont {J.}~\bibnamefont
  {Harrington}},\ and\ \bibinfo {author} {\bibfnamefont {K.}~\bibnamefont
  {Goyal}},\ }\href {https://doi.org/10.1016/j.aop.2006.01.012} {\bibfield
  {journal} {\bibinfo  {journal} {Ann. Phys.}\ }\textbf {\bibinfo {volume}
  {321}},\ \bibinfo {pages} {2242} (\bibinfo {year} {2006})}\BibitemShut
  {NoStop}%
\bibitem [{\citenamefont {Browne}\ and\ \citenamefont
  {Rudolph}(2005)}]{Browne2005-ft}%
  \BibitemOpen
  \bibfield  {author} {\bibinfo {author} {\bibfnamefont {D.~E.}\ \bibnamefont
  {Browne}}\ and\ \bibinfo {author} {\bibfnamefont {T.}~\bibnamefont
  {Rudolph}},\ }\href {https://doi.org/10.1103/PhysRevLett.95.010501}
  {\bibfield  {journal} {\bibinfo  {journal} {Phys. Rev. Lett.}\ }\textbf
  {\bibinfo {volume} {95}},\ \bibinfo {pages} {010501} (\bibinfo {year}
  {2005})}\BibitemShut {NoStop}%
\bibitem [{Note4()}]{Note4}%
  \BibitemOpen
  \bibinfo {note} {A type of time-frequency rotation could be realized with
  active operations such as chirped-pulse up-conversion~\cite {Lavoie2013-wb,
  Donohue2013-tf}.}\BibitemShut {Stop}%
\bibitem [{Note5()}]{Note5}%
  \BibitemOpen
  \bibinfo {note} {The effect of this extra $X$ gate can be eliminated by the
  way the measurement results are interpreted.}\BibitemShut {Stop}%
\bibitem [{\citenamefont {Zhang}\ \emph {et~al.}(2008)\citenamefont {Zhang},
  \citenamefont {Bao}, \citenamefont {Lu}, \citenamefont {Zhou}, \citenamefont
  {Yang}, \citenamefont {Rudolph},\ and\ \citenamefont {Pan}}]{Zhang2008-th}%
  \BibitemOpen
  \bibfield  {author} {\bibinfo {author} {\bibfnamefont {Q.}~\bibnamefont
  {Zhang}}, \bibinfo {author} {\bibfnamefont {X.-H.}\ \bibnamefont {Bao}},
  \bibinfo {author} {\bibfnamefont {C.-Y.}\ \bibnamefont {Lu}}, \bibinfo
  {author} {\bibfnamefont {X.-Q.}\ \bibnamefont {Zhou}}, \bibinfo {author}
  {\bibfnamefont {T.}~\bibnamefont {Yang}}, \bibinfo {author} {\bibfnamefont
  {T.}~\bibnamefont {Rudolph}},\ and\ \bibinfo {author} {\bibfnamefont {J.-W.}\
  \bibnamefont {Pan}},\ }\href {https://doi.org/10.1103/PhysRevA.77.062316}
  {\bibfield  {journal} {\bibinfo  {journal} {Phys. Rev. A}\ }\textbf {\bibinfo
  {volume} {77}},\ \bibinfo {pages} {062316} (\bibinfo {year}
  {2008})}\BibitemShut {NoStop}%
\bibitem [{Note6()}]{Note6}%
  \BibitemOpen
  \bibinfo {note} {This assumption would cause only quantitative differences
  since there is no need to consider factor II in the frequency
  basis.}\BibitemShut {Stop}%
\bibitem [{Note7()}]{Note7}%
  \BibitemOpen
  \bibinfo {note} {On the state preparation using time-frequency entanglement,
  the shape of coherent temporal broadening can be designed depending on the
  filter used.}\BibitemShut {Stop}%
\bibitem [{noa({\natexlab{a}})}]{noauthor_undated-nx}%
  \BibitemOpen
  \href@noop {} {\bibinfo {title} {Corning\textregistered{}
  {SMF-28e+\textregistered{}} optical fiber product information}},\ \bibinfo
  {howpublished}
  {\url{https://www.corning.com/media/worldwide/coc/documents/Fiber/PI-1463-AEN.pdf}}
  ({\natexlab{a}})\BibitemShut {NoStop}%
\bibitem [{\citenamefont {Dulkeith}\ \emph {et~al.}(2006)\citenamefont
  {Dulkeith}, \citenamefont {Xia}, \citenamefont {Schares}, \citenamefont
  {Green},\ and\ \citenamefont {Vlasov}}]{Dulkeith2006-bt}%
  \BibitemOpen
  \bibfield  {author} {\bibinfo {author} {\bibfnamefont {E.}~\bibnamefont
  {Dulkeith}}, \bibinfo {author} {\bibfnamefont {F.}~\bibnamefont {Xia}},
  \bibinfo {author} {\bibfnamefont {L.}~\bibnamefont {Schares}}, \bibinfo
  {author} {\bibfnamefont {W.~M.~J.}\ \bibnamefont {Green}},\ and\ \bibinfo
  {author} {\bibfnamefont {Y.~A.}\ \bibnamefont {Vlasov}},\ }\href
  {https://doi.org/10.1364/oe.14.003853} {\bibfield  {journal} {\bibinfo
  {journal} {Opt. Express}\ }\textbf {\bibinfo {volume} {14}},\ \bibinfo
  {pages} {3853} (\bibinfo {year} {2006})}\BibitemShut {NoStop}%
\bibitem [{\citenamefont {Korzh}\ \emph {et~al.}(2020)\citenamefont {Korzh},
  \citenamefont {Zhao}, \citenamefont {Allmaras}, \citenamefont {Frasca},
  \citenamefont {Autry}, \citenamefont {Bersin}, \citenamefont {Beyer},
  \citenamefont {Briggs}, \citenamefont {Bumble}, \citenamefont {Colangelo},
  \citenamefont {Crouch}, \citenamefont {Dane}, \citenamefont {Gerrits},
  \citenamefont {Lita}, \citenamefont {Marsili}, \citenamefont {Moody},
  \citenamefont {Pe{\~n}a}, \citenamefont {Ramirez}, \citenamefont {Rezac},
  \citenamefont {Sinclair}, \citenamefont {Stevens}, \citenamefont {Velasco},
  \citenamefont {Verma}, \citenamefont {Wollman}, \citenamefont {Xie},
  \citenamefont {Zhu}, \citenamefont {Hale}, \citenamefont {Spiropulu},
  \citenamefont {Silverman}, \citenamefont {Mirin}, \citenamefont {Nam},
  \citenamefont {Kozorezov}, \citenamefont {Shaw},\ and\ \citenamefont
  {Berggren}}]{Korzh2020-yz}%
  \BibitemOpen
  \bibfield  {author} {\bibinfo {author} {\bibfnamefont {B.}~\bibnamefont
  {Korzh}}, \bibinfo {author} {\bibfnamefont {Q.-Y.}\ \bibnamefont {Zhao}},
  \bibinfo {author} {\bibfnamefont {J.~P.}\ \bibnamefont {Allmaras}}, \bibinfo
  {author} {\bibfnamefont {S.}~\bibnamefont {Frasca}}, \bibinfo {author}
  {\bibfnamefont {T.~M.}\ \bibnamefont {Autry}}, \bibinfo {author}
  {\bibfnamefont {E.~A.}\ \bibnamefont {Bersin}}, \bibinfo {author}
  {\bibfnamefont {A.~D.}\ \bibnamefont {Beyer}}, \bibinfo {author}
  {\bibfnamefont {R.~M.}\ \bibnamefont {Briggs}}, \bibinfo {author}
  {\bibfnamefont {B.}~\bibnamefont {Bumble}}, \bibinfo {author} {\bibfnamefont
  {M.}~\bibnamefont {Colangelo}}, \bibinfo {author} {\bibfnamefont {G.~M.}\
  \bibnamefont {Crouch}}, \bibinfo {author} {\bibfnamefont {A.~E.}\
  \bibnamefont {Dane}}, \bibinfo {author} {\bibfnamefont {T.}~\bibnamefont
  {Gerrits}}, \bibinfo {author} {\bibfnamefont {A.~E.}\ \bibnamefont {Lita}},
  \bibinfo {author} {\bibfnamefont {F.}~\bibnamefont {Marsili}}, \bibinfo
  {author} {\bibfnamefont {G.}~\bibnamefont {Moody}}, \bibinfo {author}
  {\bibfnamefont {C.}~\bibnamefont {Pe{\~n}a}}, \bibinfo {author}
  {\bibfnamefont {E.}~\bibnamefont {Ramirez}}, \bibinfo {author} {\bibfnamefont
  {J.~D.}\ \bibnamefont {Rezac}}, \bibinfo {author} {\bibfnamefont
  {N.}~\bibnamefont {Sinclair}}, \bibinfo {author} {\bibfnamefont {M.~J.}\
  \bibnamefont {Stevens}}, \bibinfo {author} {\bibfnamefont {A.~E.}\
  \bibnamefont {Velasco}}, \bibinfo {author} {\bibfnamefont {V.~B.}\
  \bibnamefont {Verma}}, \bibinfo {author} {\bibfnamefont {E.~E.}\ \bibnamefont
  {Wollman}}, \bibinfo {author} {\bibfnamefont {S.}~\bibnamefont {Xie}},
  \bibinfo {author} {\bibfnamefont {D.}~\bibnamefont {Zhu}}, \bibinfo {author}
  {\bibfnamefont {P.~D.}\ \bibnamefont {Hale}}, \bibinfo {author}
  {\bibfnamefont {M.}~\bibnamefont {Spiropulu}}, \bibinfo {author}
  {\bibfnamefont {K.~L.}\ \bibnamefont {Silverman}}, \bibinfo {author}
  {\bibfnamefont {R.~P.}\ \bibnamefont {Mirin}}, \bibinfo {author}
  {\bibfnamefont {S.~W.}\ \bibnamefont {Nam}}, \bibinfo {author} {\bibfnamefont
  {A.~G.}\ \bibnamefont {Kozorezov}}, \bibinfo {author} {\bibfnamefont {M.~D.}\
  \bibnamefont {Shaw}},\ and\ \bibinfo {author} {\bibfnamefont {K.~K.}\
  \bibnamefont {Berggren}},\ }\href {https://doi.org/10.1038/s41566-020-0589-x}
  {\bibfield  {journal} {\bibinfo  {journal} {Nat. Photonics}\ }\textbf
  {\bibinfo {volume} {14}},\ \bibinfo {pages} {250} (\bibinfo {year}
  {2020})}\BibitemShut {NoStop}%
\bibitem [{Note8()}]{Note8}%
  \BibitemOpen
  \bibinfo {note} {An implementation of a phase gate induces a time delay $<
  \pi /2\omega _0 \sim 1.3$~\si {fs}, which is negligible.}\BibitemShut {Stop}%
\bibitem [{noa({\natexlab{b}})}]{noauthor_undated-rf}%
  \BibitemOpen
  \href@noop {} {\bibinfo {title} {Optoplex optical interleaver / optical
  {De-Interleaver} - symmetric interleaver and asymmetric interleaver}},\
  \bibinfo {howpublished}
  {\url{http://www.optoplex.com/Optical_Interleaver.htm}} ({\natexlab{b}}),\
  \bibinfo {note} {accessed: 2022-10-10}\BibitemShut {NoStop}%
\bibitem [{\citenamefont {Fujimoto}\ \emph {et~al.}(2022)\citenamefont
  {Fujimoto}, \citenamefont {Yamazaki}, \citenamefont {Kobayashi},
  \citenamefont {Miki}, \citenamefont {China}, \citenamefont {Terai},
  \citenamefont {Ikuta},\ and\ \citenamefont {Yamamoto}}]{Fujimoto2022-qn}%
  \BibitemOpen
  \bibfield  {author} {\bibinfo {author} {\bibfnamefont {R.}~\bibnamefont
  {Fujimoto}}, \bibinfo {author} {\bibfnamefont {T.}~\bibnamefont {Yamazaki}},
  \bibinfo {author} {\bibfnamefont {T.}~\bibnamefont {Kobayashi}}, \bibinfo
  {author} {\bibfnamefont {S.}~\bibnamefont {Miki}}, \bibinfo {author}
  {\bibfnamefont {F.}~\bibnamefont {China}}, \bibinfo {author} {\bibfnamefont
  {H.}~\bibnamefont {Terai}}, \bibinfo {author} {\bibfnamefont
  {R.}~\bibnamefont {Ikuta}},\ and\ \bibinfo {author} {\bibfnamefont
  {T.}~\bibnamefont {Yamamoto}},\ }\href {https://doi.org/10.1364/OE.469344}
  {\bibfield  {journal} {\bibinfo  {journal} {Opt. Express}\ }\textbf {\bibinfo
  {volume} {30}},\ \bibinfo {pages} {36711} (\bibinfo {year}
  {2022})}\BibitemShut {NoStop}%
\bibitem [{\citenamefont {Paesani}\ \emph {et~al.}(2021)\citenamefont
  {Paesani}, \citenamefont {Bulmer}, \citenamefont {Jones}, \citenamefont
  {Santagati},\ and\ \citenamefont {Laing}}]{Paesani2021-qn}%
  \BibitemOpen
  \bibfield  {author} {\bibinfo {author} {\bibfnamefont {S.}~\bibnamefont
  {Paesani}}, \bibinfo {author} {\bibfnamefont {J.~F.~F.}\ \bibnamefont
  {Bulmer}}, \bibinfo {author} {\bibfnamefont {A.~E.}\ \bibnamefont {Jones}},
  \bibinfo {author} {\bibfnamefont {R.}~\bibnamefont {Santagati}},\ and\
  \bibinfo {author} {\bibfnamefont {A.}~\bibnamefont {Laing}},\ }\href
  {https://doi.org/10.1103/PhysRevLett.126.230504} {\bibfield  {journal}
  {\bibinfo  {journal} {Phys. Rev. Lett.}\ }\textbf {\bibinfo {volume} {126}},\
  \bibinfo {pages} {230504} (\bibinfo {year} {2021})}\BibitemShut {NoStop}%
\bibitem [{\citenamefont {Zhang}\ \emph {et~al.}(2019)\citenamefont {Zhang},
  \citenamefont {Chen}, \citenamefont {Cui}, \citenamefont {Dowling},
  \citenamefont {Ou},\ and\ \citenamefont {Byrnes}}]{Zhang2019-dd}%
  \BibitemOpen
  \bibfield  {author} {\bibinfo {author} {\bibfnamefont {C.}~\bibnamefont
  {Zhang}}, \bibinfo {author} {\bibfnamefont {J.~F.}\ \bibnamefont {Chen}},
  \bibinfo {author} {\bibfnamefont {C.}~\bibnamefont {Cui}}, \bibinfo {author}
  {\bibfnamefont {J.~P.}\ \bibnamefont {Dowling}}, \bibinfo {author}
  {\bibfnamefont {Z.~Y.}\ \bibnamefont {Ou}},\ and\ \bibinfo {author}
  {\bibfnamefont {T.}~\bibnamefont {Byrnes}},\ }\href
  {https://doi.org/10.1103/PhysRevA.100.032330} {\bibfield  {journal} {\bibinfo
   {journal} {Phys. Rev. A}\ }\textbf {\bibinfo {volume} {100}},\ \bibinfo
  {pages} {032330} (\bibinfo {year} {2019})}\BibitemShut {NoStop}%
\bibitem [{\citenamefont {Luo}\ \emph {et~al.}(2019)\citenamefont {Luo},
  \citenamefont {Zhong}, \citenamefont {Erhard}, \citenamefont {Wang},
  \citenamefont {Peng}, \citenamefont {Krenn}, \citenamefont {Jiang},
  \citenamefont {Li}, \citenamefont {Liu}, \citenamefont {Lu}, \citenamefont
  {Zeilinger},\ and\ \citenamefont {Pan}}]{Luo2019-hx}%
  \BibitemOpen
  \bibfield  {author} {\bibinfo {author} {\bibfnamefont {Y.-H.}\ \bibnamefont
  {Luo}}, \bibinfo {author} {\bibfnamefont {H.-S.}\ \bibnamefont {Zhong}},
  \bibinfo {author} {\bibfnamefont {M.}~\bibnamefont {Erhard}}, \bibinfo
  {author} {\bibfnamefont {X.-L.}\ \bibnamefont {Wang}}, \bibinfo {author}
  {\bibfnamefont {L.-C.}\ \bibnamefont {Peng}}, \bibinfo {author}
  {\bibfnamefont {M.}~\bibnamefont {Krenn}}, \bibinfo {author} {\bibfnamefont
  {X.}~\bibnamefont {Jiang}}, \bibinfo {author} {\bibfnamefont
  {L.}~\bibnamefont {Li}}, \bibinfo {author} {\bibfnamefont {N.-L.}\
  \bibnamefont {Liu}}, \bibinfo {author} {\bibfnamefont {C.-Y.}\ \bibnamefont
  {Lu}}, \bibinfo {author} {\bibfnamefont {A.}~\bibnamefont {Zeilinger}},\ and\
  \bibinfo {author} {\bibfnamefont {J.-W.}\ \bibnamefont {Pan}},\ }\href
  {https://doi.org/10.1103/PhysRevLett.123.070505} {\bibfield  {journal}
  {\bibinfo  {journal} {Phys. Rev. Lett.}\ }\textbf {\bibinfo {volume} {123}},\
  \bibinfo {pages} {070505} (\bibinfo {year} {2019})}\BibitemShut {NoStop}%
\bibitem [{\citenamefont {Dai}\ \emph {et~al.}(2013)\citenamefont {Dai},
  \citenamefont {Liu}, \citenamefont {Gao}, \citenamefont {Xu},\ and\
  \citenamefont {He}}]{Dai2013-uv}%
  \BibitemOpen
  \bibfield  {author} {\bibinfo {author} {\bibfnamefont {D.}~\bibnamefont
  {Dai}}, \bibinfo {author} {\bibfnamefont {L.}~\bibnamefont {Liu}}, \bibinfo
  {author} {\bibfnamefont {S.}~\bibnamefont {Gao}}, \bibinfo {author}
  {\bibfnamefont {D.-X.}\ \bibnamefont {Xu}},\ and\ \bibinfo {author}
  {\bibfnamefont {S.}~\bibnamefont {He}},\ }\href
  {https://doi.org/10.1002/lpor.201200023} {\bibfield  {journal} {\bibinfo
  {journal} {Laser Photon. Rev.}\ }\textbf {\bibinfo {volume} {7}},\ \bibinfo
  {pages} {303} (\bibinfo {year} {2013})}\BibitemShut {NoStop}%
\bibitem [{\citenamefont {Azuma}\ \emph {et~al.}(2015)\citenamefont {Azuma},
  \citenamefont {Tamaki},\ and\ \citenamefont {Lo}}]{Azuma2015-fr}%
  \BibitemOpen
  \bibfield  {author} {\bibinfo {author} {\bibfnamefont {K.}~\bibnamefont
  {Azuma}}, \bibinfo {author} {\bibfnamefont {K.}~\bibnamefont {Tamaki}},\ and\
  \bibinfo {author} {\bibfnamefont {H.-K.}\ \bibnamefont {Lo}},\ }\href
  {https://doi.org/10.1038/ncomms7787} {\bibfield  {journal} {\bibinfo
  {journal} {Nat. Commun.}\ }\textbf {\bibinfo {volume} {6}},\ \bibinfo {pages}
  {6787} (\bibinfo {year} {2015})}\BibitemShut {NoStop}%
\bibitem [{\citenamefont {Ewert}\ \emph {et~al.}(2016)\citenamefont {Ewert},
  \citenamefont {Bergmann},\ and\ \citenamefont {van Loock}}]{Ewert2016-ge}%
  \BibitemOpen
  \bibfield  {author} {\bibinfo {author} {\bibfnamefont {F.}~\bibnamefont
  {Ewert}}, \bibinfo {author} {\bibfnamefont {M.}~\bibnamefont {Bergmann}},\
  and\ \bibinfo {author} {\bibfnamefont {P.}~\bibnamefont {van Loock}},\ }\href
  {https://doi.org/10.1103/PhysRevLett.117.210501} {\bibfield  {journal}
  {\bibinfo  {journal} {Phys. Rev. Lett.}\ }\textbf {\bibinfo {volume} {117}},\
  \bibinfo {pages} {210501} (\bibinfo {year} {2016})}\BibitemShut {NoStop}%
\bibitem [{\citenamefont {Borregaard}\ \emph {et~al.}(2020)\citenamefont
  {Borregaard}, \citenamefont {Pichler}, \citenamefont {Schr{\"o}der},
  \citenamefont {Lukin}, \citenamefont {Lodahl},\ and\ \citenamefont
  {S{\o}rensen}}]{Borregaard2020-jk}%
  \BibitemOpen
  \bibfield  {author} {\bibinfo {author} {\bibfnamefont {J.}~\bibnamefont
  {Borregaard}}, \bibinfo {author} {\bibfnamefont {H.}~\bibnamefont {Pichler}},
  \bibinfo {author} {\bibfnamefont {T.}~\bibnamefont {Schr{\"o}der}}, \bibinfo
  {author} {\bibfnamefont {M.~D.}\ \bibnamefont {Lukin}}, \bibinfo {author}
  {\bibfnamefont {P.}~\bibnamefont {Lodahl}},\ and\ \bibinfo {author}
  {\bibfnamefont {A.~S.}\ \bibnamefont {S{\o}rensen}},\ }\href
  {https://doi.org/10.1103/PhysRevX.10.021071} {\bibfield  {journal} {\bibinfo
  {journal} {Phys. Rev. X}\ }\textbf {\bibinfo {volume} {10}},\ \bibinfo
  {pages} {021071} (\bibinfo {year} {2020})}\BibitemShut {NoStop}%
\bibitem [{\citenamefont {Kuhlmann}\ \emph {et~al.}(2015)\citenamefont
  {Kuhlmann}, \citenamefont {Prechtel}, \citenamefont {Houel}, \citenamefont
  {Ludwig}, \citenamefont {Reuter}, \citenamefont {Wieck},\ and\ \citenamefont
  {Warburton}}]{Kuhlmann2015-zs}%
  \BibitemOpen
  \bibfield  {author} {\bibinfo {author} {\bibfnamefont {A.~V.}\ \bibnamefont
  {Kuhlmann}}, \bibinfo {author} {\bibfnamefont {J.~H.}\ \bibnamefont
  {Prechtel}}, \bibinfo {author} {\bibfnamefont {J.}~\bibnamefont {Houel}},
  \bibinfo {author} {\bibfnamefont {A.}~\bibnamefont {Ludwig}}, \bibinfo
  {author} {\bibfnamefont {D.}~\bibnamefont {Reuter}}, \bibinfo {author}
  {\bibfnamefont {A.~D.}\ \bibnamefont {Wieck}},\ and\ \bibinfo {author}
  {\bibfnamefont {R.~J.}\ \bibnamefont {Warburton}},\ }\href
  {https://doi.org/10.1038/ncomms9204} {\bibfield  {journal} {\bibinfo
  {journal} {Nat. Commun.}\ }\textbf {\bibinfo {volume} {6}},\ \bibinfo {pages}
  {8204} (\bibinfo {year} {2015})}\BibitemShut {NoStop}%
\bibitem [{\citenamefont {Lavoie}\ \emph {et~al.}(2013)\citenamefont {Lavoie},
  \citenamefont {Donohue}, \citenamefont {Wright}, \citenamefont {Fedrizzi},\
  and\ \citenamefont {Resch}}]{Lavoie2013-wb}%
  \BibitemOpen
  \bibfield  {author} {\bibinfo {author} {\bibfnamefont {J.}~\bibnamefont
  {Lavoie}}, \bibinfo {author} {\bibfnamefont {J.~M.}\ \bibnamefont {Donohue}},
  \bibinfo {author} {\bibfnamefont {L.~G.}\ \bibnamefont {Wright}}, \bibinfo
  {author} {\bibfnamefont {A.}~\bibnamefont {Fedrizzi}},\ and\ \bibinfo
  {author} {\bibfnamefont {K.~J.}\ \bibnamefont {Resch}},\ }\href
  {https://doi.org/10.1038/nphoton.2013.47} {\bibfield  {journal} {\bibinfo
  {journal} {Nat. Photonics}\ }\textbf {\bibinfo {volume} {7}},\ \bibinfo
  {pages} {363} (\bibinfo {year} {2013})}\BibitemShut {NoStop}%
\bibitem [{\citenamefont {Donohue}\ \emph {et~al.}(2013)\citenamefont
  {Donohue}, \citenamefont {Agnew}, \citenamefont {Lavoie},\ and\ \citenamefont
  {Resch}}]{Donohue2013-tf}%
  \BibitemOpen
  \bibfield  {author} {\bibinfo {author} {\bibfnamefont {J.~M.}\ \bibnamefont
  {Donohue}}, \bibinfo {author} {\bibfnamefont {M.}~\bibnamefont {Agnew}},
  \bibinfo {author} {\bibfnamefont {J.}~\bibnamefont {Lavoie}},\ and\ \bibinfo
  {author} {\bibfnamefont {K.~J.}\ \bibnamefont {Resch}},\ }\href
  {https://doi.org/10.1103/PhysRevLett.111.153602} {\bibfield  {journal}
  {\bibinfo  {journal} {Phys. Rev. Lett.}\ }\textbf {\bibinfo {volume} {111}},\
  \bibinfo {pages} {153602} (\bibinfo {year} {2013})}\BibitemShut {NoStop}%
\end{thebibliography}%


\begin{thebibliography}{0}%
\makeatletter
\providecommand \@ifxundefined [1]{%
 \@ifx{#1\undefined}
}%
\providecommand \@ifnum [1]{%
 \ifnum #1\expandafter \@firstoftwo
 \else \expandafter \@secondoftwo
 \fi
}%
\providecommand \@ifx [1]{%
 \ifx #1\expandafter \@firstoftwo
 \else \expandafter \@secondoftwo
 \fi
}%
\providecommand \natexlab [1]{#1}%
\providecommand \enquote  [1]{``#1''}%
\providecommand \bibnamefont  [1]{#1}%
\providecommand \bibfnamefont [1]{#1}%
\providecommand \citenamefont [1]{#1}%
\providecommand \href@noop [0]{\@secondoftwo}%
\providecommand \href [0]{\begingroup \@sanitize@url \@href}%
\providecommand \@href[1]{\@@startlink{#1}\@@href}%
\providecommand \@@href[1]{\endgroup#1\@@endlink}%
\providecommand \@sanitize@url [0]{\catcode `\\12\catcode `\$12\catcode
  `\&12\catcode `\#12\catcode `\^12\catcode `\_12\catcode `\%12\relax}%
\providecommand \@@startlink[1]{}%
\providecommand \@@endlink[0]{}%
\providecommand \url  [0]{\begingroup\@sanitize@url \@url }%
\providecommand \@url [1]{\endgroup\@href {#1}{\urlprefix }}%
\providecommand \urlprefix  [0]{URL }%
\providecommand \Eprint [0]{\href }%
\providecommand \doibase [0]{https://doi.org/}%
\providecommand \selectlanguage [0]{\@gobble}%
\providecommand \bibinfo  [0]{\@secondoftwo}%
\providecommand \bibfield  [0]{\@secondoftwo}%
\providecommand \translation [1]{[#1]}%
\providecommand \BibitemOpen [0]{}%
\providecommand \bibitemStop [0]{}%
\providecommand \bibitemNoStop [0]{.\EOS\space}%
\providecommand \EOS [0]{\spacefactor3000\relax}%
\providecommand \BibitemShut  [1]{\csname bibitem#1\endcsname}%
\let\auto@bib@innerbib\@empty
\end{thebibliography}%
\end{document}


\section{Supplemental material}
\subsection{A. Derivations of Equations~\eqref{wavepacket} and \eqref{physical_qudit_temp}}
Herein, we list useful formulas and derive Eqs.~\eqref{wavepacket} and \eqref{physical_qudit_temp} in a more general form that includes the effect of group velocity dispersion as a function $D$.

Note that the following relations hold.
\begin{align}
    T_{\omega}(f\odot g)&=T_{\omega}(f) \odot T_{\omega}(g), 
    \label{relation1} \\
    T_{\omega}(f*g)&=T_{\omega}(f)*g=f*T_{\omega}(g), 
    \label{relation2} \\
    M_{\tau}(f \odot g)&=M_{\tau}(f) \odot g=f \odot M_{\tau}(g), 
    \label{relation3} \\
    M_{\tau}(f*g)&=M_{\tau}(f)*M_{\tau}(g),
    \label{relation4}
\end{align}
\begin{equation}\label{relation_Fourier}
    \hat{(T_{\omega}(f))}=M_{-\omega}(\hat{f}), \quad \hat{(M_{\tau}(f))}=T_{\tau}(\hat{f})
\end{equation}
and 
\begin{equation}\label{comm_relation}
    T_{\omega} M_{\tau} = e^{i\omega \tau}M_{\tau} T_{\omega}
\end{equation}
for functions $f$ and $g$.

A photon wave packet can be converted between the time and frequency domains as follows,
\begin{equation}\label{creation}
    (\xi * a^\dag)(0) = (\hat{\xi} * \hat{a}^\dag)(0).
\end{equation}
It holds that 
\begin{equation}
    (T_{\omega_0}(D\odot M_{\tau_0}(\xi)) * a^\dag)(0)
    =  (M_{-\omega_0}(\hat{D}*T_{\tau_0}(\hat{\xi})) * \hat{a}^\dag)(0)
    = e^{i\omega_0 \tau_0}(T_{\tau_0}(M_{-\omega_0}(\hat{D}*\hat{\xi}))*\hat{a}^\dag)(0),
\end{equation}
where the first equality holds from Eqs.~\eqref{creation} and \eqref{relation_Fourier}, and the second equality holds from Eqs.~\eqref{relation2} and \eqref{comm_relation}. 
Thus, we have 
\begin{equation}\label{wavepacket_D}
    ((D\odot M_{\tau_0}(\xi)) * a^\dag)(\omega_0)\ket*{0} = e^{i\omega_0\tau_0}(M_{-\omega_0}(\hat{D}*\hat{\xi})*\hat{a}^\dag)(\tau_0)\ket*{0}.
\end{equation}
We obtain Eq.~\eqref{wavepacket} by omitting $D$ from Eq.~\eqref{wavepacket_D}.

Using the Poisson summation formula for a Dirac comb,
\begin{equation}
    C_{\overline{x}}(x)=\sum_n\delta(x-n\overline{x})=\frac{1}{\overline{x}}\sum_n e^{i2\pi n \frac{x}
    {\overline{x}}},
\end{equation}
it is derived that
\begin{equation}\label{discrete_continuous}
    \sum_{k=0}^{d-1} e^{-i2\pi jk/d}T_{k\overline{x}/d}(C_{\overline{x}})= M_{2\pi j/\overline{x}}(C_{\overline{x}/d}).
\end{equation}
The Fourier transformation of a Dirac comb is also a Dirac comb, as follows:
\begin{equation}\label{comb_Fourier}
    \hat{C}_{\omega_r} = \frac{2\pi}{\omega_r}C_{2\pi/\omega_r}.
\end{equation}

The subsequent state following the propagation of a physical time--frequency Gottesmann--Kitaev--Preskill (TFGKP) state with central frequency $\omega_0$ for time $\tau_0$ is described as
\begin{equation}
    \ket*{\tilde{i}_f} \propto ((D \odot M_{\tau_0}(\phi_t) \odot (T_{i\omega_r/d}(C_{\omega_r})*\phi_f))*a^\dag) (\omega_0) \ket*{0}
\end{equation}
Correspondingly, it holds that
\begin{equation}
    \ket*{\tilde{j}_t} \propto \sum_{k=0}^{d-1} e^{-i2\pi jk/d} \ket*{\tilde{k}_f} 
    =((D\odot M_{\tau_0}(\phi_t) \odot (M_{2\pi j/\omega_r}(C_{\omega_r/d})*\phi_f))*a^\dag)(\omega_0)\ket*{0}
\end{equation}
from Eq.~\eqref{discrete_continuous}.
The Fourier transformation of $D'=D \odot (M_{2\pi j/\omega_r}(C_{\omega_r/d})*\phi_f)$ is
\begin{equation}
    \hat{D'} = \hat{D}*(T_{j\tau_r/d}(C_{\tau_r}) \odot \hat{\phi_f})
\end{equation}
from Eqs.~\eqref{relation_Fourier} and \eqref{comb_Fourier}.
Therefore, by replacing $D$ by $D'$ in Eq.~\eqref{wavepacket_D}, we obtain
\begin{equation}\label{temp_GKP_formula}
    \ket*{\tilde{j}_t} \propto e^{i\omega_0 \tau_0}(M_{-\omega_0}(\hat{D}*( T_{j\tau_r/d}(C_{\tau_r}) \odot \hat{\phi_f})*\hat{\phi_t})*\hat{a}^\dag)(\tau_0)\ket*{0}.
\end{equation}
Omitting $D$ from Eq.~\eqref{temp_GKP_formula} yields Eq.~\eqref{physical_qudit_temp}.

\subsection{B. Detailed error analysis}
Herein, we derive formulas for the error probabilities.
We explicitly represent $\ket*{\tilde{i}_f}$ with central frequency $\omega_0$ as $\ket*{\tilde{i}_f(\omega_0)}$ and $\ket*{\tilde{i}_t}$ with central timing $\tau_0$ as $\ket*{\tilde{i}_t(\tau_0)}$.
We use PAFs $\phi_{t/f}$ for coherent temporal/spectral broadening and PDFs $\Phi_{t/f}$ for incoherent temporal/spectral broadening.
For all the calculations listed below, we assume that temporal/spectral broadenings of TFGKP states are sufficiently narrow that the overlaps between the bins are negligible, except within half of the time/frequency periods.

The error probability of qudit measurement in the temporal basis with a time-resolving detector is, 
\begin{equation}\label{Et1}
    e_{t,1} = 1-\sum_n\int_{-\frac{\tau_r}{2d}}^{\frac{\tau_r}{2d}} d\tau \int_{-\infty}^{\infty} d\tau' \Phi_t(\tau') 
    = \abs{\bra{0}\hat{a}(\tau_0 + \tau +n \tau_r)\ket*{\tilde{0}_t(\tau_0-\tau')}}^2.
\end{equation}
We have
\begin{align}
    \abs{\bra{0}\hat{a}(\tau_0 + \tau +n \tau_r)\ket*{\tilde{0}_t(\tau_0-\tau')}}^2 
     &= \frac{1}{N} \abs{((\hat{\phi}_f \odot C_{\tau_r})*\hat{\phi}_t)(-(\tau'+\tau+n\tau_r))}^2 \\
     &\simeq \frac{1}{\sum_m \abs*{\hat{\phi}_f(m\tau_r)}^2} \abs*{\hat{\phi}_f(-n\tau_r) \hat{\phi}_t(-(\tau'+\tau))}^2,
\end{align}
with a normalization factor $N\simeq \sum_m \abs*{\hat{\phi}_f(m\tau_r)}^2$.
Therefore,
\begin{equation}\label{Et1_res}
    e_{t,1} \simeq 1-\int_{-\frac{\tau_r}{2d}}^{\frac{\tau_r}{2d}} d\tau (\Phi_t*(\hat{\phi}_t^* \odot \hat{\phi}_t))(\tau)
\end{equation}
holds.
Similarly, we express the error probability of qudit measurement in the frequency basis with a frequency-resolving detector as,
\begin{equation}\label{Ef1_res}
    e_{f,1} \simeq 1-\int_{-\frac{\omega_r}{2d}}^{\frac{\omega_r}{2d}} d\omega (\Phi_f*(\phi_f^* \odot \phi_f))(\omega).
\end{equation}

When we use the $1$:$d$ optical interleavers (OIs) and $d$ detectors instead of a frequency-resolving detector for measurement in the frequency basis, the error probability becomes
\begin{equation}\label{Ef1'}
    e_{f,1} = \frac{\sum_{k=1}^{d-1}F_k}{\sum_{k=0}^{d-1} F_k} \simeq \frac{F_1+F_{d-1}}{F_0}
\end{equation}
where
\begin{equation}
    F_k = \int_{-\infty}^{\infty} d\omega \int_{-\infty}^{\infty} d\omega' \Phi_f(\omega')
    \abs{\bra{0}(T_{-\omega'}(I_k) \odot a)(\omega) \ket*{\tilde{0}_f(\omega_0)}}^2.
\end{equation}
Assuming good OI, that is, $T_{\omega_r}(g_I) \simeq g_I$ and $f_I(\omega) \simeq 0$ for $\abs{\omega} > \omega_r/2$,
\begin{align}
    F_x &\simeq \frac{1}{N} \int_{-\infty}^{\infty} d\omega \int_{-\infty}^{\infty} d\omega' \Phi_f(\omega')
    \abs{(T_{\omega'}(g_I) \odot \phi_t \odot (C_{\omega_r}*(T_{x\omega_r/d+\omega'}(f_I) \odot \phi_f)))(-\omega)}^2 \\
    &\simeq \frac{1}{N} \sum_n\abs{(g_I \odot \phi_t)(n\omega_r)}^2 
    \int_{-\infty}^{\infty}d\omega (T_{x\omega_r/d}(f_I^* \odot f_I) \odot (\Phi_f * (\phi_f^* \odot \phi_f)))(\omega),
\end{align}
where $x=-1,0,1$ and $F_{-1} = F_{d-1}$.
Therefore, $e_{f,1}$ hardly depends on the envelope of the OI $g_I$.
When $f_I(\omega)=1$ for $\abs{\omega} \leq \omega_r/2d$ and $f_I(\omega)=0$ for $\abs{\omega} > \omega_r/2d$, Eq.~\eqref{Ef1'} becomes the same as Eq.~\eqref{Ef1_res}.
We assume this for simplicity.

In the generations of entangled states and measurement on entangled bases, partial distinguishability causes a deviation of the states from the maximally entangled states.
The temporal distinguishability between the two input photons is, 
\begin{equation}\label{Et2_res}
    e_{t,2} = 1- \int_{-\infty}^{\infty}d \tau \Phi_t(\tau)\abs{\bra*{\tilde{0}_t(\tau_0)}\ket*{\tilde{0}_t(\tau_0-\tau)}}^2 
    \simeq 1-  (\Phi_t*G_t)(0),
\end{equation}
where
\begin{equation}
    G_t(\delta)= \abs{\int_{-\infty}^{\infty} d\tau \hat{\phi}_t^*(\tau) \hat{\phi}_t(\tau+\delta)}^2.
\end{equation}
In principle, this formula should be generalized to multiphoton interference, such as four-photon interference in bell-state generations.
Herein, we use Eq.~\eqref{Et2_res} as the error probability to characterize the effect of temporal distinguishability.

Assuming that the concrete shapes of coherent/incoherent temporal/spectral broadenings, we calculated the error probabilities.
We assumed that the incoherent temporal broadening was mainly caused by the finite temporal resolution of the detectors used for the generation and measurement of a qudit, which is usually characterized by a Gaussian function $\Phi_{t}=f_{G,\sigma_{t,i}}$, where 
\begin{equation}
    f_{G,\sigma}(x) = \frac{1}{(2 \pi \sigma^2)^{1/2}} \exp(-\frac{x^2}{2\sigma^2}).
\end{equation}
We assume that the coherent temporal broadening is mainly determined by the state-generation method based on which we can design its PAF using the appropriate filter during state generation.
Thus, we consider $\phi_{t}^* \odot \phi_{t}$ to be Gaussian, that is, $\phi_{t}=(8\pi \sigma_{t,c}^2)^{1/4} f_{G,\sqrt{2}\sigma_{t,c}}$.
We also assumed that the coherent spectral broadening is determined by the state generation method, but it is reasonable to assume that $\phi_f^* \odot \phi_f$ is Lorentzian, that is, $\phi_f = f_{L,\gamma_{f,c}}$ for
\begin{equation}
    f_{L,\gamma}(x) = \sqrt{\frac{\gamma}{\pi}}\frac{1}{\gamma-ix},
\end{equation}
because this is the case for the transmission spectrum of the cavities.
For example, incoherent broadening may be induced by the instability of OIs.
However, we ignored this for simplicity because incoherent spectral broadening plays the same role as coherent spectral broadening in our scheme.

Eqs.~\eqref{Et1_res}, \eqref{Ef1_res}, and \eqref{Et2_res} correspond to the following equations,
\begin{align}
    e_{t,1} &\simeq 1-\text{erf}\qty\Big(\frac{\tau_r}{2d}(\sigma_{t,c}^2+\sigma_{t,i}^2)^{-\frac{1}{2}}) \\
    e_{f,1} &\simeq 1-\frac{2}{\pi}\text{Tan}^{-1}\qty\Big(\frac{\omega_r/2d}{\gamma_{f,c}}) \\
    e_{t,2} &\simeq 1- \qty\Big(1+ \frac{\sigma_{t,i}^2}{2\sigma_{t,c}^2})^{-1/2}.
\end{align}
We can translate them into the expression with the FWHMs $\Delta_\text{FWHM}$ of each PDF, where $\Delta_\text{FWHM}=\sqrt{8\ln{2}}\sigma$ for a Gaussian and $\Delta_\text{FWHM}=2\gamma$ for a Lorentzian.
Setting their common error threshold $e$, we finally set the following inequalities as a sufficient condition:
\begin{align}
    \frac{\Delta_{t,i}}{\Delta_{t,c}} & \lesssim \sqrt{A} \\
    \frac{\Delta_{t,c}}{\tau_r/d} & \lesssim \sqrt{\frac{2\ln{2}}{1+A}}(\text{erf}^{-1}(1-e))^{-1} \\
    \frac{\Delta_{f,c}}{\omega_r/d} & \lesssim \tan(\frac{\pi}{2}(1-e))^{-1},
\end{align}
where $A = 2((1-e)^{-2}-1)$, and $\Delta_{t,i}$, $\Delta_{t,c}$, and $\Delta_{f,c}$ are the full-width-at-half-maximum (FWHM) values of the incoherent temporal, coherent temporal, and coherent spectral broadenings, respectively.
We obtain Eq.~\eqref{errorthereshold}, for $e=0.01$.

\makeatletter\@input{tfgkp_aux.tex}\makeatother